\documentclass[a4paper,fleqn,times]{cas-sc}
\usepackage{array}
\usepackage{amsfonts}
\usepackage{bm}
\usepackage{amssymb}
\usepackage{amsmath}
\usepackage{lineno}
\usepackage{mathtools}
\usepackage{xcolor}
\usepackage{graphicx}
\usepackage{dcolumn}
\usepackage{booktabs}
\usepackage{multicol}
\usepackage{multirow}
\usepackage{rccol}
\usepackage{tabularx}
\usepackage{suffix}
\usepackage{lipsum}
\usepackage{siunitx}
\usepackage{comment}
\usepackage{csquotes}
\usepackage{enumitem}
\usepackage{datetime}
\usepackage[noabbrev]{cleveref}
\usepackage[sort&compress,longnamesfirst,numbers]{natbib}
\newcommand{\vectorsym}[1]{\boldsymbol{#1}}

\newcommand{\tensorsym}[1]{\boldsymbol{#1}}

\newcommand{\diff}[1]{\,\mathrm{d}#1}
\newcommand{\dn}[2]{\dfrac{\diff#1}{\diff#2}}

\newcommand{\dpn}[2]{\dfrac{\partial#1}{\partial#2}}

\newcolumntype{C}{>{\centering\arraybackslash}X}
\newcolumntype{N}[2]{>{\centering\arraybackslash}R[.][.]{#1}{#2}}
\newcommand{\figurespath}{./}
\ExplSyntaxOn
\keys_set:nn { stm / mktitle } { nologo }
\ExplSyntaxOff
\begin{document}
%
%
\def\floatpagepagefraction{1}
\def\textpagefraction{0.001}
\shorttitle{Implementation of IBVP Methods in CODA}
\shortauthors{Núñez et al.}
\title[mode=title]{Implementation of Immersed Boundaries via Volume Penalization in the Industrial Aeronautical Computational Fluid Dynamics Solver CODA}
\author[1,2]{Jonatan Núñez}[%
  auid=001,%
  bioid=001,%
  orcid=0000-0002-4541-2768]
\cormark[1]
\ead{jonatan.nunez@upm.es}
\credit{Conceptualization of this study, Methodology, Software}
\author[1,2]{David Huergo}[%
  auid=002,%
  bioid=002,%
  orcid=0009-0008-9091-5824]
\ead{david.huergo.perea@upm.es}
\credit{Conceptualization of this study, Methodology, Software}
\author[1,3]{Diego Lodares}[%
  auid=003,%
  bioid=003,%
  orcid=0000-0002-9841-0803]
\ead{diego.lodares@airbus.com}
\credit{Conceptualization of this study, Methodology, Software}
\author[1,2]{Suyash Shrestha}[%
  auid=004,%
  bioid=004,%
  orcid=0009-0000-5760-7012]
\ead{s.shrestha@upm.es}
\credit{Conceptualization of this study, Methodology, Software}
\author[3]{Juan Guerra}[%
  auid=005,%
  bioid=005,%
  ]
\credit{Conceptualization of this study, Methodology, Software}
\author[3]{Juan Florenciano}[%
  auid=006,%
  bioid=006,%
  orcid=0000-0002-8189-6982]
\ead{juan.florenciano@airbus.com}
\credit{Conceptualization of this study, Methodology, Software}
\author[1,2]{Esteban Ferrer}[%
  auid=007,%
  bioid=007,%
  orcid=0000-0003-1519-0444]
\ead{esteban.ferrer@upm.es}
\credit{Conceptualization of this study, Methodology, Software, Funding}
\author[1,2]{Eusebio Valero}[%
  auid=008,%
  bioid=008,%
  orcid=0000-0002-1627-6883]
\ead{eusebio.valero@upm.es}
\credit{Conceptualization of this study, Methodology, Software, Funding}
\affiliation[1]{%
  organization={School of Aeronautics, Universidad Politécnica de Madrid},
  addressline={Plaza del Cardenal Cisneros 3},
  city={Madrid},
  postcode={28040},
  country={Spain}%
}
\affiliation[2]{%
  organization={Center for Computational Simulation, Universidad Politécnica de Madrid},
  addressline={Campus Montegancedo},
  city={Boadilla del Monte},
  postcode={28660},
  country={Spain}%
}
\affiliation[3]{%
  organization={Airbus Defence and Space S.A.U – Flight Physics},
  addressline={Paseo John Lennon, s/n},
  city={Getafe},
  postcode={28906},
  country={Spain}%
}
\cortext[1]{Corresponding author}
%
\begin{abstract}
  We present the implementation and validation of an immersed boundary volume penalization method in the computational fluid dynamics solver CODA (from ONERA, DLR, and Airbus). Our goal is to model and simulate turbulent fluid flows in complex 3D aerodynamic configurations through the numerical solution of the Reynolds--averaged Navier--Stokes equations using the Spalart--Allmaras turbulent model. To do that, an immersed boundary method has been implemented in CODA and an efficient preprocessing tool for the construction of unstructured hexahedral meshes with adaptive mesh refinement around immersed geometries has been developed.\par
  We report several numerical examples, including subsonic flow past the NACA0012 airfoil, transonic flow past the RAE2822 airfoil, subsonic flow past the MDA30P30N multi-element airfoil, and subsonic flow around the NASA high-lift CRM aircraft. These simulations have been performed in the CODA solver with a second-order finite volume scheme as spatial discretization and an implicit backward Euler scheme based on the matrix-free GMRES block-Jacobi iterative method. The reported numerical simulations are in good agreement with their corresponding experimental data. These encouraging results allow us to conclude that the implemented immersed boundary method is efficient, flexible, and accurate and can therefore be used for aeronautical applications in industry.
\end{abstract}
%
%
%
%
\maketitle
%
%
\section{Introduction}
In the ever-evolving landscape of computational fluid dynamics (CFD) and industrial simulations, the accurate representation of complex geometries and their interaction with fluid flow remains a critical challenge. The ability to model and analyze intricate structures in a computationally efficient manner is paramount, as it directly influences the design, performance, and optimization of various industrial systems. This challenge is particularly relevant in the context of the industrial solver CODA \cite{volpiani2024a}, where the interaction between fluid flow and complex geometries plays a central role.\par
CODA is a parallel software framework for multidisciplinary analysis and optimization of aircraft and helicopters based on advanced and accurate numerical methods \cite{kroll2016a,jaegerskuepper2022a}. The solver is being developed as part of a collaboration between the French National Aerospace Research Center (ONERA), the German Aerospace Center (DLR) and Airbus. CODA provides a robust, scalable and computational efficient integrated design process for aerodynamics and structural analysis. To perform an efficient analysis and optimization of aircraft on state-of-the-art HPC systems, the Navier--Stokes and the Reynolds-averaged Navier--Stokes (RANS) equations are solved for high Reynolds-number flow on unstructured grids with second-order finite-volume and higher-order discontinuous-Galerkin discretizations.\par
The implementation of immersed boundaries through volume penalization has emerged as a promising technique to address the difficulty to generate body fitted meshes for complex geometries and simulate moving geometries. Immersed boundaries allow for the inclusion of detailed and intricate geometries within the computational domain, even in cases where traditional structured grids might be impractical or prohibitively expensive to generate. In the realm of CFD simulations, the term \enquote{immersed boundary} refers to the representation of complex structures, such as solid objects or porous media, as part of the fluid computational domain, without the need for a grid that conforms to their intricate shapes. This not only enhances the capability to model real-world industrial scenarios, but also significantly improves the computational efficiency (while reducing human intervention) of the simulation workflow because the time-consuming body-fitted mesh generation is avoided.\par
The primary objective of this research is to provide an in-depth exploration of the methodology, mathematical formulations, and computational considerations involved in the implementation of immersed boundaries through volume penalization in the solver CODA. By doing so, this study aims to facilitate a deeper understanding of this technique and its potential applications in aeronautics. It is our hope that the insights and findings presented in this paper will contribute to the advancement of CFD simulations, offering engineers and researchers a valuable tool for the accurate modeling of complex industrial systems and the optimization of their performance.\par
The immersed boundary method was introduced by Peskin \cite{peskin1972a} for analyzing the flow through the native mitral heart valve, and since then, the method has been applied to the numerical simulation of flow problems over complex geometries \cite{iaccarino2003a}, multiphase flows \cite{shao2013a}, and fluid-structure interaction \cite{wang2015d}.\par
The main feature of immersed boundary methods is that the governing equations are solved on a Cartesian background mesh which is not necessarily conforming with the immersed body. Immersed boundary methods can be classified according to the mechanisms used to describe the interaction between the fluid and the immersed geometry. Several approaches are currently available, among them, the more relevant are the cut-cell methods \cite{ye1999a,udaykumar2001a,orley2015a}, the ghost cell methods \cite{tseng2003a,hu2013c,xia2014a,khalili2018b}, the direct forcing approach \cite{goldstein1993a,fadlun2000a,luo2012a,boukharfane2018a}, sharp interface methods \cite{ghias2007a,de-vanna2020a}, and the volume penalization schemes \cite{angot1999a,liu2007a,schneider2015a,engels2015a,abalakin2016a}. In \citep{kou2022a} is discussed the efficiency of the immersed boundary volume penalization for handling moving geometries, like the flow around an airfoil with pitching and plunging motions. More details on different families of immersed boundary methods can be found in the reviews \cite{mittal2005a,sotiropoulos2014a,griffith2020a,verzicco2023a} and the references cited therein. Recently, immersed boundary methods have been extensively used to simulate turbulent aerodynamic flow problems in the context of RANS simulations \cite{capizzano2007a,capizzano2011a,capizzano2018a,constant2021a,capizzano2016a,peron2017a,renaud2019a,peron2021a}.
This paper is organized as follows: In \cref{sec:numerical-methods} we summarize the main numerical schemes implemented in CODA for simulating turbulent fluid flow with immersed boundary methods. In \cref{sec:preprocessing-postprocessing} we discuss the preprocessing and postprocessing tools developed for generating refined meshes around immersed bodies and the analysis of simulation data. Next, in \cref{sec:numerical-computations} we present several numerical computations done with CODA with immersed boundary volume penalization methods: subsonic flow past the NACA0012 airfoil, transonic flow past the RAE2822 airfoil, subsonic flow past the MDA30P30N multi-element airfoil and a preliminary simulation of the subsonic flow around the NASA high-lift CRM aircraft. Finally, in \cref{sec:conclusions}, a summary of this work is presented, and advantages and disadvantages of the immersed boundary volume penalization for aerodynamic simulations in industrial scenarios are discussed.
%
%
\section{Governing Equations}\label{sec:equations}
\subsection{Navier--Stokes Equations}\label{sec:equations:nse}
The Navier--Stokes equations can be written as a hyperbolic-parabolic system of conservation laws in the following way
\begin{equation*}
  \dpn{\vectorsym{u}}{t}+\nabla\cdot\vectorsym{f}(\vectorsym{u},\nabla{\vectorsym{u}})=\vectorsym{0}.
\end{equation*}
The vector of conserved quantities is defined by $\vectorsym{u}(\vectorsym{x},t)=\left(\rho,\rho\vectorsym{v},\rho{}E\right)$, where $\rho$ is the mass density, $\vectorsym{v}=\left(v_{x},v_{y},v_{z}\right)$ is the velocity vector and $E$ is the total energy. The physical flux is defined by
\begin{equation*}
  \vectorsym{f}(\vectorsym{u},\nabla\vectorsym{u})=
    \vectorsym{f}^{A}(\vectorsym{u})-\vectorsym{f}^{D}(\vectorsym{u},\nabla\vectorsym{u}),
\end{equation*}
with the convective and viscous fluxes given, respectively, by
\begin{equation*}
  \vectorsym{f}^{A}(\vectorsym{u})=%
    \begin{pmatrix}
      \rho\,\vectorsym{v}\\
      \rho\,\vectorsym{v}\otimes\vectorsym{v}+p\tensorsym{I}\\
      \vectorsym{v}\left(\rho{}E+p\right)
    \end{pmatrix},\quad
  \vectorsym{f}^{D}(\vectorsym{u},\nabla\vectorsym{u})=%
    \begin{pmatrix}
      0\\
      -\tensorsym{\tau}\\
      \tensorsym{\tau}\cdot\vectorsym{v}-\vectorsym{q}
    \end{pmatrix}.
\end{equation*}
The viscous stresses are described by the stress tensor $\tensorsym{\tau}$, defined by
\begin{equation*}
  \tensorsym{\tau}=2\mu{}\tensorsym{S}^{D}=\mu\left(\nabla\vectorsym{v}+\left(\nabla\vectorsym{v}\right)^{\intercal}-\frac{2}{3}\left(\nabla\cdot\vectorsym{v}\right)\tensorsym{I}\right),
\end{equation*}
where $\tensorsym{S}^{D}$ is the deviatoric component of the strain-rate tensor
\begin{equation*}
  \tensorsym{S}=\frac{1}{2}\left(\nabla\vectorsym{v}+\left(\nabla\vectorsym{v}\right)^{\intercal}\right)
\end{equation*}
and
\begin{equation*}
  \vectorsym{q}=-k\nabla{T},\quad k=\mu\frac{c_{p}}{\mathrm{Pr}}.
\end{equation*}
The equation of state
\begin{equation*}
  p=\left(\gamma-1\right)\left(\rho{}E-\frac{1}{2}\rho{}v^{2}\right),
\end{equation*}
where $p$ is the static pressure, $\gamma$ is the ideal gas index, $\mu$ is the dynamic viscosity, $T$ is the temperature, $k$ is the thermal conductivity, and $\mu$ denotes the dynamic viscosity coefficient. The kinematic viscosity coefficient is defined by the formula
\begin{equation*}
  \nu = \mu/\rho.
\end{equation*}
\subsection{Reynolds-Averaged Navier--Stokes Equations}\label{sec:equations:rans}
The Navier--Stokes equations govern the motion of fluids in both laminar and turbulent regimes. Because of the large range of spatial and temporal scales present in turbulent flows, directly solving the Navier--Stokes equations is prohibitively expensive. Therefore, the Reynolds--averaged Navier--Stokes (RANS) equations are solved instead to model steady turbulent mean flows. The RANS equations couple the mean flow equations with the one-equation turbulence model of Spallart--Allmaras. These equations can be written also as a hyperbolic-parabolic system of conservation laws with source term in the following way
\begin{equation*}
  \dpn{\vectorsym{u}}{t}+\nabla\cdot\vectorsym{f}(\vectorsym{u},\nabla{\vectorsym{u}})=\vectorsym{s}\left(\vectorsym{u}\right).
\end{equation*}
Here, $\vectorsym{u}$ is the vector of time-averaged conservative variables over a given time interval and has the following components: $\vectorsym{u}(\vectorsym{x},t)=\left(\rho,\rho\vectorsym{v},\rho{}E,\rho\tilde{\nu}\right)$, where $\rho$ is the time-averaged mass density, $\vectorsym{v}=\left(v_{x},v_{y},v_{z}\right)$ is the time-averaged velocity vector, $E$ is the time-averaged total energy, and $\rho\tilde{\nu}$ is a new conservative variable relating the time-averaged mass density and the eddy viscosity $\tilde{\nu}$. The advective and diffusive components of the physical flux associated with the quantity $\rho\tilde{\nu}$ are given by
\begin{equation*}
  \vectorsym{f}^{A}\left[\rho\tilde{\nu}\right]=\rho\tilde{\nu}\vectorsym{v},\quad
  \vectorsym{f}^{D}\left[\rho\tilde{\nu}\right]=\frac{1}{\sigma}\left(\mu+f_{n1}\rho\tilde{\nu}\right)\nabla{}\tilde{\nu}
\end{equation*}
In the diffusive fluxes, the turbulent stress tensor $\tensorsym{\tau}_{t}$ and the turbulent
heat fluxes $\vectorsym{q}_{t}$ are added respectively to $\tensorsym{\tau}$ and $\vectorsym{q}$ appearing in the mean flow equations:
\begin{equation*}
  \tensorsym{\tau}_{t}=2\mu_{t}\tensorsym{S}^{D},\quad
  \vectorsym{q}_{t}=-\frac{\mu_{t}}{\mathrm{Pr}_{t}}c_{p}\nabla{}T
\end{equation*}
where $\mathrm{Pr}_{t}=0.9$ is the turbulent Prandtl number and $\mu_{t}$ is the turbulent dynamic viscosity
\begin{equation*}
  \mu_{t}=
    \begin{cases}
      \rho\tilde{\nu}f_{v1}\left(\chi\right) & \text{for $\tilde{\nu}\ge{}0$},\\
      0 & \text{for $\tilde{\nu}<0$},\\
    \end{cases},\quad
  f_{v1}\left(\chi\right)=\frac{\chi^{3}}{\chi^{3}+c_{v1}^{3}},\quad
  \chi=\frac{\rho\tilde{\nu}}{\mu}.
\end{equation*}
The source terms only have nonzero component in the equation for the turbulent variable $\rho\tilde{\nu}$:
\begin{equation*}
  \vectorsym{S}\left[\rho\tilde{\nu}\right]=-\rho\left(P-D\right)-\frac{c_{b2}}{\sigma}\rho\nabla\tilde{\nu}+\frac{1}{\sigma}\left(\nu+f_{n1}\tilde{\nu}\right)\nabla\rho\cdot\nabla\tilde{\nu},
\end{equation*}
where the production and destruction terms, $P$ and $D$, are defined by:
\begin{align*}
  P&=
    \begin{cases}
      c_{b1}\left(1-f_{t2}\right)\tilde{\omega}\tilde{\nu} & \text{for $\tilde{\nu}\ge{}0$},\\
      c_{b1}\left(1-c_{t3}\right)\omega\tilde{\nu} & \text{for $\tilde{\nu}<0$},\\
    \end{cases}\\
  D&=
    \begin{cases}
      \left(c_{w1}f_{w}-\frac{c_{b1}}{\kappa^{2}}f_{t2}\right)\left(\frac{\tilde{\nu}}{d}\right)^{2}
       & \text{for $\tilde{\nu}\ge{}0$},\\
      -c_{w1}\left(\frac{\tilde{\nu}}{d}\right)^{2} & \text{for $\tilde{\nu}<0$},\\
    \end{cases}
\end{align*}
and
\begin{equation*}
  f_{n1}=
    \begin{cases}
      1 & \text{for $\tilde{\nu}\ge{}0$},\\
      \frac{c_{n1}+\chi^{3}}{c_{n1}-\chi^{3}} & \text{for $\tilde{\nu}<0$},\\
    \end{cases},\quad
  f_{t2}=c_{t3}\exp{\left(-c_{t4}\chi^{2}\right)},\quad f_{w}=g\left(\frac{1+c_{w3}^{6}}{g^{6}+c_{w3}^{6}}\right)^{\frac{1}{6}}
\end{equation*}
with
\begin{equation*}
  g=r+c_{w2}\left(r^{6}-r\right),\quad r=\min\left(r_{\mathrm{lim}},\frac{\tilde{\nu}}{\omega\kappa^{2}d^{2}}\right),
\end{equation*}
and $d$ is the distance to the nearest wall and $\omega$ the vorticity magnitude. The modified vorticity magnitude $\tilde{\omega}$ is given by
\begin{equation*}
  \tilde{\omega}=
    \begin{cases}
      \omega+\bar{\omega} & \text{for $\bar{\omega}>-c_{v2}\omega$},\\
      \omega+\frac{\omega\left( c_{v2}^{2}+c_{v3}\bar{\omega}\right)}{\left(c_{v3}-2c_{v2}\right)\omega-\bar{\omega}} & \text{for $\bar{\omega}<-c_{v2}\omega$},
    \end{cases}
\end{equation*}
where $\bar{\omega}$ and $f_{v2}$ are given by
\begin{equation*}
  \bar{\omega}=\frac{\tilde{\nu}}{\kappa^{2}d^{2}}f_{v2},\quad f_{v2}=1-\frac{\chi}{1-f_{v1}}
\end{equation*}
For the sake of completeness, we give the values of the constants in the above expressions:
$c_{v1}=\num{7.1}$,
$\sigma=2/3$,
$c_{b1}=\num{0.1355}$,
$c_{b2}=\num{0.622}$,
$\kappa=\num{0.41}$,
$c_{w2}=\num{0.3}$,
$c_{w3}=2$,
$r_{\mathrm{lim}}=\num{10}$,
$c_{t3}=1.2$,
$c_{t4}=0.5$,
$c_{v2}=0.7$,
$c_{v3}=0.9$,
$c_{n1}=16$.
%
%
\section{Numerical Methods}\label{sec:numerical-methods}
CODA is a CFD solver, which  integrates an extensive set of algorithms for solving partial differential equations in a multi-physics context. In this work, we focus on the implementation of an immersed boundary methodology for the modelling of turbulent fluid flows by the well-known Reynolds-Averaged Navier--Stokes equations. In our formulation, we have selected the Spalart--Allmaras turbulent model that we briefly reproduce in \cref{sec:equations} for completeness of the paper. Regarding the spatial and time discretization employed in CODA, we summarize the key ingredients in the next section.
\subsection{Spatial and Time Discretization}\label{sec:numerical-methods:space-time}
The spatial discretization schemes used in CODA to solve the Navier--Stokes equations and the Reynolds-Averaged Navier--Stokes equations are the finite volume method \cite{blazek2015a}, the modal discontinuous Galerkin method \cite{hesthaven2008a}, and the discontinuous Galerkin spectral element method \cite{kopriva2009a,giraldo2020a}. The partial differential equations of interest can be written as a hyperbolic-parabolic system of conservation laws in differential form (see \cref{sec:equations} for more details)
\begin{equation*}
  \dpn{\vectorsym{u}}{t}+\nabla\cdot\vectorsym{f}(\vectorsym{u},\nabla{\vectorsym{u}})=\vectorsym{s}\left(\vectorsym{u}\right),
\end{equation*}
where $\vectorsym{u}=\vectorsym{u}(\vectorsym{x},t)$ is the vector of conserved quantities, and $\vectorsym{f}=\vectorsym{f}(\vectorsym{u},\nabla\vectorsym{u})$ is the tensor of physical fluxes, which has two terms, the advective flux and the diffusive term, namely
\begin{equation*}
  \vectorsym{f}(\vectorsym{u},\nabla\vectorsym{u})=
    \vectorsym{f}^{A}(\vectorsym{u})-\vectorsym{f}^{D}(\vectorsym{u},\nabla\vectorsym{u}),
\end{equation*}
and $\vectorsym{s}=\vectorsym{s}\left(\vectorsym{u}\right)$ represents the source terms. The conservation laws can also be written in integral form as follows
\begin{equation*}
  \dpn{}{t}\int_{\Omega}\vectorsym{u}\diff{\Omega} +
  \oint_{\partial\Omega}\left(\vectorsym{f}^{A}(\vectorsym{u})-\vectorsym{f}^{D}(\vectorsym{u},\nabla\vectorsym{u})\right)\cdot\vectorsym{n}\diff{\sigma} = \int_{\Omega}\vectorsym{s}\diff{\Omega}.
\end{equation*}
Here, $\Omega$ is the spatial domain and $\vectorsym{n}$ is the normal vector on the surface $\partial\Omega$ enclosing the spatial domain. In this work, we are interested only in the finite-volume schemes. A short description of these methods is outlined; for the discontinuous Galerkin schemes, see \cite{basile2022a}. In the finite-volume framework, the computational domain is divided into a set of non-overlapping polyhedral control volumes or cells, that is, $\Omega=\bigcup_{i}^{N}\Omega_{i}$, and the integral form of the equations is discretized for each control volume $\Omega_{i}$.
Defining the mean value of the function $\vectorsym{u}$ in the cell $\Omega_{i}$ by
\begin{equation*}
  \vectorsym{u}_{i}:=\dfrac{1}{\left|\Omega_{i}\right|}\int_{\Omega_{i}}\vectorsym{u}\left(\vectorsym{x},t\right)\diff{\Omega},
\end{equation*}
we get the so-called semi-discrete formulation
\begin{equation*}
  \dn{\vectorsym{u}_{i}}{t}:=-\dfrac{1}{\left|\Omega_{i}\right|}
    \biggl\{
    \sum_{l}\bigl[\vectorsym{f}^{A}\cdot\hat{\vectorsym{n}}\bigr]^{*}\sigma_{l}-\sum_{l}\bigl[\vectorsym{f}^{D}\cdot\hat{\vectorsym{n}}\bigr]^{*}\sigma_{l}
    - \vectorsym{s}\left|\Omega_{i}\right|
    \biggr\}.
\end{equation*}
The surface integral has been approximated by a sum of the fluxes through the faces $\partial\Omega_{l}$ of the cell $\Omega_{i}$. The $\sigma_{l}$ stands for the area of the face $\partial\Omega_{l}$, $\hat{\vectorsym{n}}$ is the normal vector to the face $\partial\Omega_{l}$, and $\left|\Omega_{i}\right|$ is the volume of the cell $\Omega_{i}$. The flux across the element faces is computed via numerical fluxes (this is represented in the equation as the operator $\left[\cdot\right]^{*}$). The computation of the numerical advective fluxes requires a reconstruction procedure that takes cell-centered values of conserved quantities, momentum and temperature gradients, and interpolate them to the faces. Typically, the Roe schemes with an entropy fix are employed. The gradients are computed based on the Green--Gauss theorem. This procedure reconstructs values needed at face integration points to approximate the Green--Gauss integral linearly. Gradient limiters can be applied in the vecinity of strong discontinuities. The finite volume discretization implemented in CODA is second-order accurate. More details on how are computed the element and face gradients can be found in \cite{schwoeppe2013a,volpiani2024a,langer2024a}.\par
Regarding the time discretization, in CODA are available explicit, implicit and implicit-explicit integrators. In this work we have used an implicit backward-Euler scheme based on preconditioned matrix-free GMRES.
Preconditioners include incomplete LU and line-inversion \cite{mohnke2023a}.
\subsection{Immersed Boundary Methods}\label{sec:numerical-methods:ibm}
Immersed Boundary Methods are numerical techniques for simulating fluid-structure interaction problems involving complex geometries, providing an alternative to numerical methods based on body-fitted meshes. Immersed Boundary Methods were first proposed by \citet{peskin1972a} to simulate blood flow in the human heart. These methods provide a flexible framework for modeling fluid flow over complicated immersed bodies on grids that do not conform to the surface of the body by treating these as immersed boundaries within the computational domain. In this way, immersed boundary methods overcome the challenges of generating high-quality meshes around complex geometries, which can be time-consuming and computationally demanding, while accurately resolving complex flows using simple grids. The geometric components are discretized using a separate representation, such as a surface or volume mesh, which is then immersed within the computational grid. Immersed boundary conditions are applied to take into account the geometric components lying in the flow.\par
Because the meshes used in the immersed boundary method are, in principle, non-conforming with the geometry of the obstacle, a very high resolution of this background mesh is required to obtain reliable numerical solutions. The employment of very fine meshes in the whole computational domain makes unpractical the use of immersed boundary methods for industrial applications, like the study of the aerodynamics of aircrafts. In fact, very fine meshes are only necessary in specific places in the computational domain, like around the immersed geometry and the wake regions. Fine meshes far away from the immersed bodies are unnecessary. For these reasons, mesh refinement algorithms are used to refine the mesh only in the regions where it is actually essential for accurate computations.
\subsubsection{Immersed Boundary Volume Penalization Method}\label{sec:numerical-methods:ibvp}
\begin{figure}[t]
  \centering
  \includegraphics[width=0.49\linewidth]{\figurespath/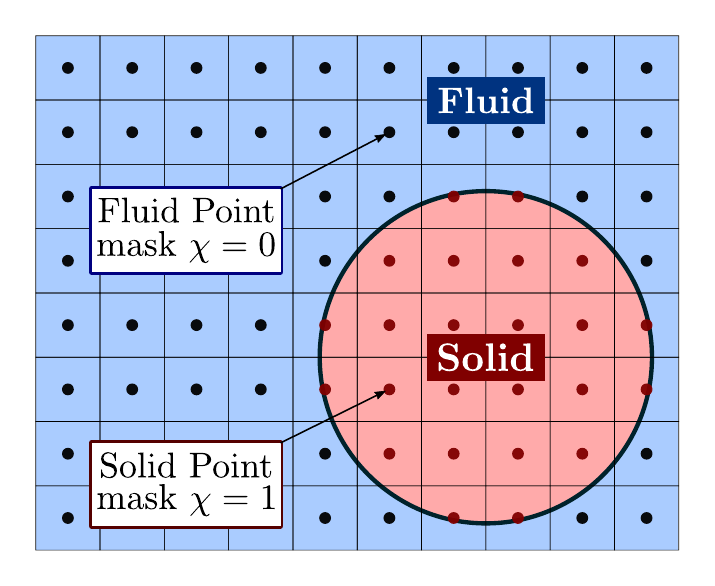}%
  \hfil%
  \includegraphics[width=0.49\linewidth]{\figurespath/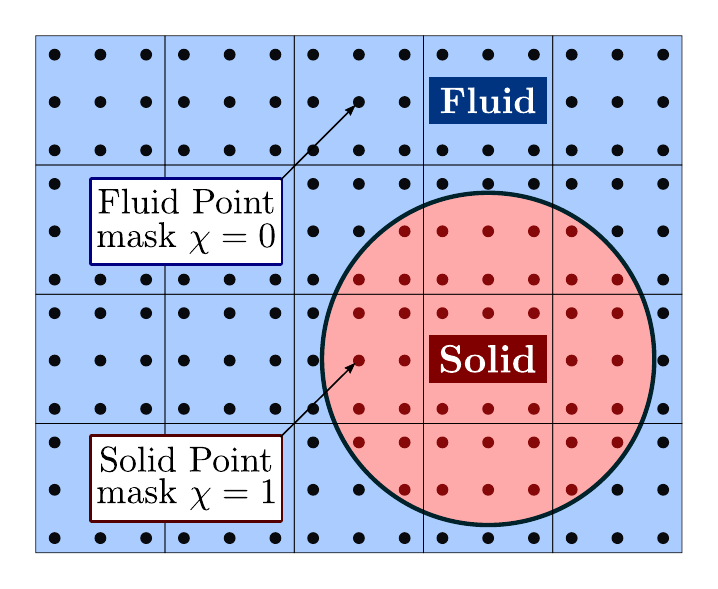}%
  \caption{Schematic diagram of the background mesh for the immersed boundary method as it used  along with finite volume schemes (left) and discontinuous Galerkin spectral element methods (right).}%
  \label{fig:mesh:ibvp}%
\end{figure}

In this family of immersed boundary volume penalization methods, the flow equations are solved on the whole computational domain, while a source term is introduced to represent the body as a porous medium with very low permeability
\cite{kou2022a,kou2022b,llorente2023a,llorente2024a,ferrer2023a}. This method is based on the idea of penalizing the integration points which are located inside the immersed body using source terms, instead of using a boundary condition on the surface of the body, as it is the case when using conforming meshes. In this way, a simple Cartesian mesh can be used to solve the Navier--Stokes equations over arbitrary geometries, reducing the time required for the mesh generation \cref{fig:mesh:ibvp}.\par
\begin{figure}[t]
  \centering
  \includegraphics[width=0.75\linewidth]{\figurespath/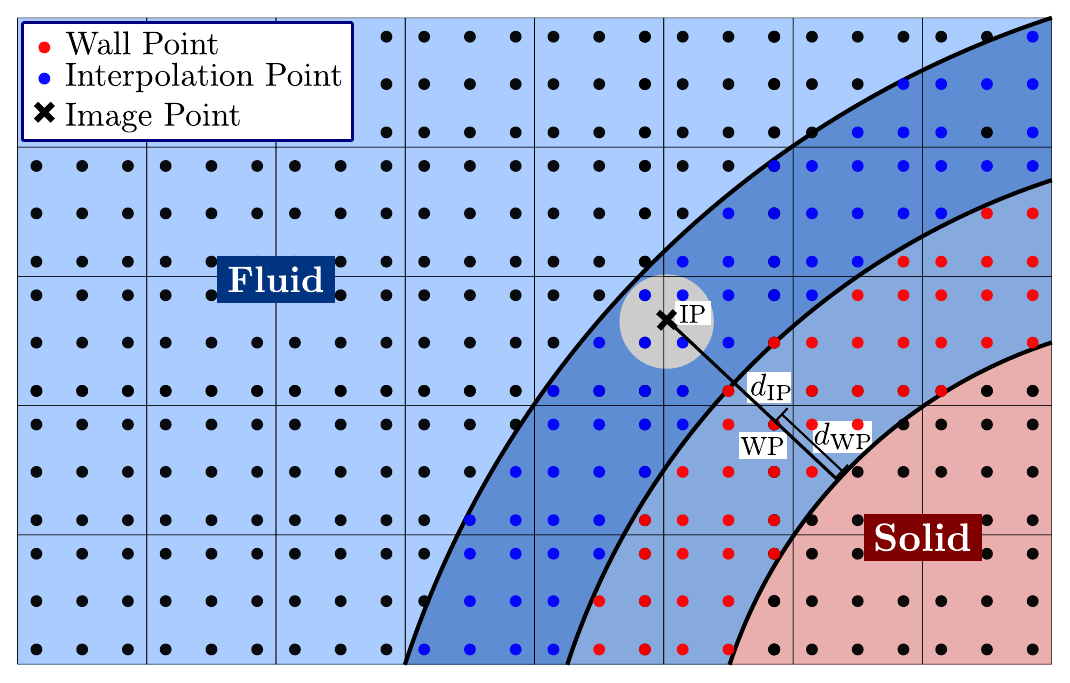}%
  \caption{Schematic definition of wall points, image points and interpolation points.}%
  \label{fig:mesh:ibvp:wall-points}%
\end{figure}

In the IBVP method, a mask function is required to discriminate between those points that are inside or outside the immersed body, as it is represented in \cref{fig:mesh:ibvp}. Then, source terms are applied to modify the behavior of the flow field inside the immersed body. Given the governing equations for a compressible viscous fluid:
\begin{equation}
  \dpn{\vectorsym{u}}{t}+\nabla\cdot\vectorsym{f}(\vectorsym{u},\nabla{\vectorsym{u}})=\vectorsym{s}(\vectorsym{u})
\end{equation}
where $\vectorsym{s}(\vectorsym{u})$ is the IBM source term. This source term is written as
\begin{equation}
  \vectorsym{s}(\vectorsym{u};\chi,\eta)=%
  \frac{\chi}{\eta}
    \begin{pmatrix}
      0\\
      \rho(\vectorsym{v}-\vectorsym{v}_{\mathrm{s}})\\
      \dfrac{1}{2}\rho{}\left(\vectorsym{v}\cdot\vectorsym{v}-\vectorsym{v}_{\mathrm{s}}\cdot\vectorsym{v}_{\mathrm{s}}\right)
    \end{pmatrix},
\end{equation}
where $\vectorsym{v}_{\mathrm{s}}$ is the velocity of the moving geometry (note that for static object $\vectorsym{v}_{\mathrm{s}}=0$ and$\chi$ represents the mask function and distinguishes between the fluid region $\Omega_{\mathrm{f}}$ and the solid region $\Omega_{\mathrm{s}}$:
\begin{equation}
  \chi\left(\vectorsym{x},t\right) =%
  \begin{cases}
    1, & \text{if $\vectorsym{x}\in\Omega_{\mathrm{s}}$},\\
    0, & \text{otherwise}.
  \end{cases}
\end{equation}
and $0<\eta\ll{}1$ is the penalization parameter. If a RANS Spalart--Allmaras model is used, the eddy viscosity, $\tilde{\nu}$, is set to zero inside the body \cite{tamaki2017a}:
\begin{equation}
  \vectorsym{s}_{\mathrm{SA}}(\vectorsym{u};\chi,\eta)=-\frac{\chi}{\eta}\tilde{\nu}.
\end{equation}
Furthermore, a wall model has been developed to improve the accuracy of RANS simulations. This model selects a set of points near the immersed body where the wall function will be applied, which are called wall points. For each wall point, the normal vector to the body's surface is computed and a new point is defined, called image point. The image points will provide the required information to the wall function to update the state vector at the wall points. However, image points (in general) are not integration points of the computational mesh. Therefore, the state vector at the image points has to be interpolated from a set of interpolation points. These interpolation points, which are computed through a kd-tree algorithm, are the $N$-nearest integration points to each image point. An overview of this approach is represented schematically in \cref{fig:mesh:ibvp:wall-points}. The distance which is used to discriminate between the regions where the wall points and the image points live, is defined as the length of the smallest edge of the smallest cell of the mesh.\par
To interpolate the state vector at each image point, a Radial Basis Function (RBF) interpolator with a Gaussian kernel is used. Then, a wall function is used to compute the friction velocity at each image point \cite{frere2017a}:
\begin{equation}
  v_{||}^{+}(y^{+})=\frac{1}{\kappa}\log{\left(1+\kappa{}y^{+}\right)} + \left(C-\frac{1}{\kappa}\log{(\kappa)}\right)\left(1-\exp\left(-\frac{y^{+}}{11}\right)-\frac{y^{+}}{11}\exp\left(-\frac{y^{+}}{3}\right)\right),
\end{equation}
and
\begin{equation}
  v_{\mathrm{IP},t} = v_{\tau}v_{||}^{+}\left(y^{+}_{\mathrm{IP}}\right),\quad\text{with}\quad
    y^{+}_{\mathrm{IP}} = \frac{v_{\tau}d_{\mathrm{IP}}}{\nu_{\mathrm{IP}}},
\end{equation}
where the subscript $\mathrm{IP}$ refers to each image point and the subscript $t$ refers to the tangential component of the velocity vector.\par
Finally, the resulting friction velocity $v_{\tau}$ is considered to be constant along the normal direction and, hence, it can be used at the wall points (subscript $\mathrm{WP}$). Taking into account this information, the velocity, the eddy viscosity and the non-dimensional wall distance can be obtained at the wall points:
\begin{align}
  y^{+}_{\mathrm{WP}} &= \frac{v_{\tau}d_{\mathrm{WP}}}{\nu_{\mathrm{WP}}}, \\
  v_{\mathrm{WP},t}   &= v_{\tau}v^{+}\left(y^{+}_{\mathrm{WP}}\right), \\
  v_{\mathrm{WP},n}   &= \frac{d_{\mathrm{WP}}}{d_{\mathrm{IP}}}v_{\mathrm{IP},n}, \\
  \tilde{\nu}_{\mathrm{WP}} &= \kappa{}v_{\tau}d_{\mathrm{WP}},
\end{align}
which can be solved including an additional condition of parallelism:
\begin{equation}
  v_{\mathrm{WP},t} \parallel v_{\mathrm{IP},t}
\end{equation}
The subscript $n$ refers to the normal direction of the velocity vector, $\kappa=0.41$ is the von Kármán constant and $d$ is the distance from a point to the immersed body. The new velocity and turbulent viscosity at the wall points are considered by including additional source terms for the momentum, energy and turbulent viscosity equations. Finally, this wall model as implemented in CODA along with the IBVP scheme has proved to improve the results of the standard RANS simulation if the $y^{+}$ at the wall points has a value (approximately) of $60$ or less.
%
%
\section{Preprocessing and Postprocessing}\label{sec:preprocessing-postprocessing}
\subsection{Preprocessing}\label{sec:preprocessing}
Immersed boundary methods use a background mesh for solving a fluid-structure interaction problem. In 3D applications, this background mesh is typically a simple Cartesian mesh made of hexahedral elements. However, it can be also unstructured and made of hybrid elements (hexahedra, prisms, tetrahedra, etc.), or even a body-fitted mesh. In this section we focus on the construction of a Cartesian background mesh built out of hexahedra. This kind of mesh is assembled in an unstructured fashion by the preprocessing tool.
\subsubsection{Background Mesh Generation}\label{sec:preprocessing:mesh-generation}
We have developed a preprocessing tool to create meshes made of hexahedral elements. The preprocessing tool has two ways for generating the initial background mesh: the first way consists in creating a new Cartesian mesh from scratch, and the second way consists in importing a mesh generated by the software GMSH \cite{geuzaine2009a} (strictly speaking, the preprocessing tool is not generating a mesh but importing an already generated one by an external tool). Our preprocessing tool is only currently capable of handling meshes made of hexahedral elements. Next, we will describe the tasks the preprocessing tool does for constructing an unstructured and adaptively refined Cartesian mesh.
\subsubsection*{Loading Parameters for Background Mesh Generation}
The first task to be performed is to load from file the parameters of the mesh to be constructed. Depending on the type of mesh to be generated (built-in or imported mesh), different parameters are read-in.\par
For the built-in mesh type, the preprocessing tool requires some parameters that specify the type of built-in box (standard box, curved box, box deformation function, type of target geometry, mesh stretching factors in each spatial direction, etc.). The number of elements in each direction of the Cartesian box, the degree of the mesh (for high-order curved meshes), and the kind of curved boundary are also required.\par
For the imported mesh type, the filename of the GMSH file is required. Additional parameters are necessary for setting up the name and location of boundary conditions and the output format of the mesh file (formats supported: HDF5, GMSH, and TECPLOT).
\subsubsection*{Creation of the Background Mesh}
Once the mesh parameters are loaded, the preprocessing tool proceeds to create the background mesh, either a built-in or an imported mesh. The typical workflow of the preprocessing tool is as follows:
\begin{itemize}
  \item {Creation of interpolation matrices between different polynomial bases.}
  \item {Creation of a reference box.}
  \item {Creation of a curved box from the reference box.}
  \item {Application of the stretching functions on each spatial direction, if required.}
  \item {Application of mesh deformation functions, if required.}
  \item {Application of mesh rotation matrices, if required.}
  \item {Storing the final mesh in arrays.}
  \item {Numbering of nodes, elements, elements faces and boundary faces.}
  \item {Creation of the main data structure of the mesh: the list of elements with their corresponding information (nodes ID, elements ID, local faces definition, boundary conditions). This elements list is a dynamical data structure, namely, a linked list. Taking into account that the number of elements increases dynamically according to the desired target mesh size around the immersed geometry, this data structure makes possible a straightforward application of the mesh refinement algorithm.}
\end{itemize}
\subsubsection{Mesh Refinement around Immersed Geometries}\label{sec:preprocessing:mesh-refinement}
Once the background mesh is generated and stored in the element list data structure, the preprocessing tool will refine this mesh provided the geometry of the body is in STL format.
\subsubsection*{Loading Parameters for the Mesh Refinement}
From the same parameter file used in the generation of the background mesh, the preprocessing tool will load all the information necessary for the mesh refinement task: the maximum level of refinement around the immersed geometry, the number and location of special regions that need to be refined, and their corresponding levels of refinement, the type of refinement (either isotropic or anisotropic), and the filenames of the STL files used to describe the geometry of the immersed bodies. After reading this information, the preprocessing tool will perform mesh refinement around the immersed geometry.
\subsubsection*{Importing Geometries from STL Files}
The immersed geometries are described by STL files. These files contain an unstructured triangulated surface (triangle unit vectors and the coordinates of their corresponding vertices) as generated by the CAD software. The preprocessing tool is responsible for loading the tessellated surface of the immersed body for further use in the refinement stage.
\subsubsection*{Mesh Refinement around Immersed Geometries}
The preprocessing tool has the capability of refining the background hexahedral mesh in special regions within the computational domain (for instance, wake regions) and also around the triangulated surface of the immersed geometry. The preprocessing tool will follow the next steps for refining the background mesh:
\begin{itemize}
  \item {\textbf{Flagging the elements that require refinement}. For the refinement of special regions in the computational domain, the flagging procedure marks all hexahedral cells inside the region coordinates provided in the parameter file. For the refinement of the hexahedral elements intersecting the triangulated surface of the immersed body, an efficient algorithm for the triangle-box overlap is implemented \cite{akenine2001a}. Before mesh refinement takes place, this triangle-box overlap is tested for all triangles of the surface triangulation, and if such overlap takes place, the involved elements store the overlapped triangles. During the mesh refinement loop (refinement level $l>0$), this triangle-box overlap is tested only for those hexahedra storing overlapped triangles.}
  \item {\textbf{Flagging the elements for balancing}. The FSDM and, therefore, CODA can handle hanging nodes for different types of face elements (triangular and quadrilateral faces). This hanging nodes capability is restricted in FSDM to elements with a $\text{2:1}$ refinement level ratio. This means that for two neighboring elements with different refinement levels, only a level difference of at most $1$ between neighboring cells is allowed. The preprocessing tool is aware of this restriction and therefore an algorithm for preserving the $\text{2:1}$ balancing has been implemented.}
  \item {A feature implemented in the preprocessing tool is the refinement of elements around the immersed body up to certain number of neighboring cells far away from the truly overlapped elements with the geometry. This allows us to have a thick layer of small elements around the immersed body, which is useful in the simulation of flow problems with the RANS equations and the use of wall functions.}
\end{itemize}
\begin{figure}[ht]
  \centering
  \includegraphics[width=0.45\linewidth]{\figurespath/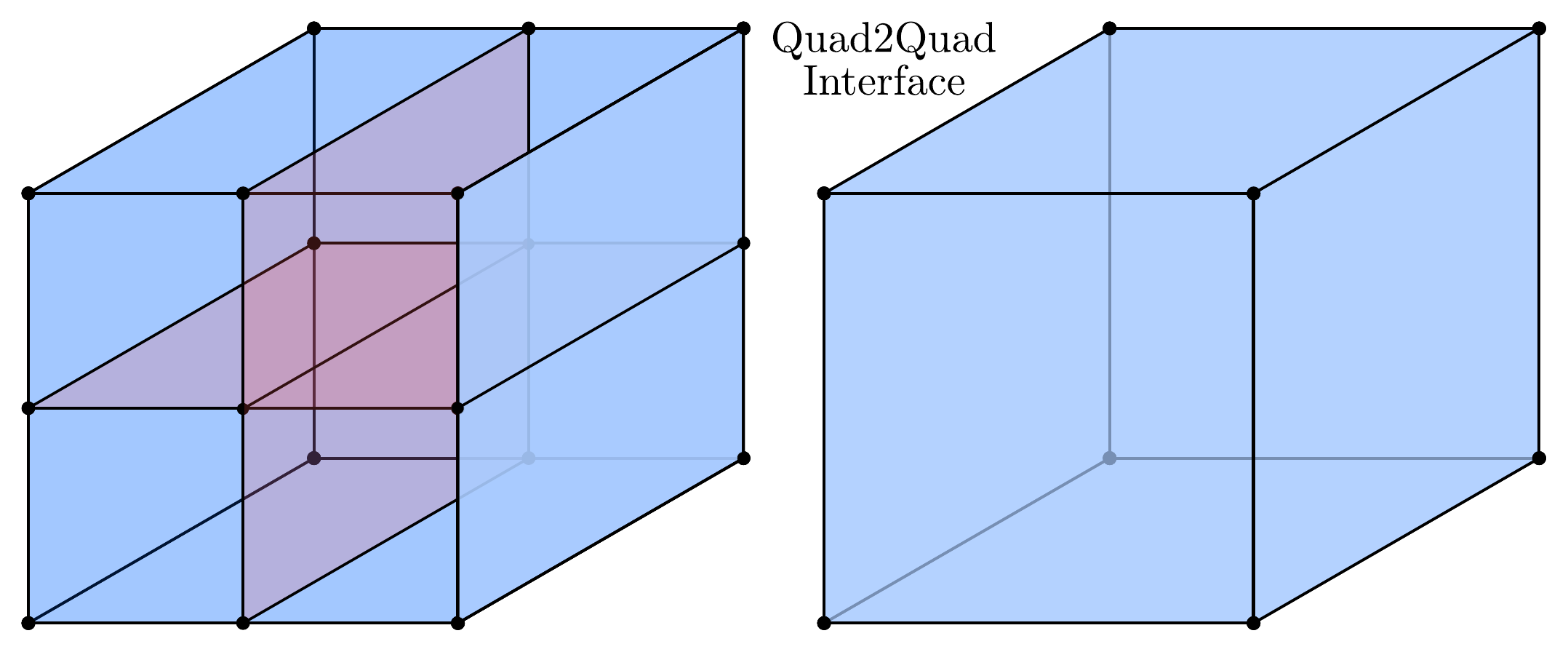}%
  \hfil
  \includegraphics[width=0.45\linewidth]{\figurespath/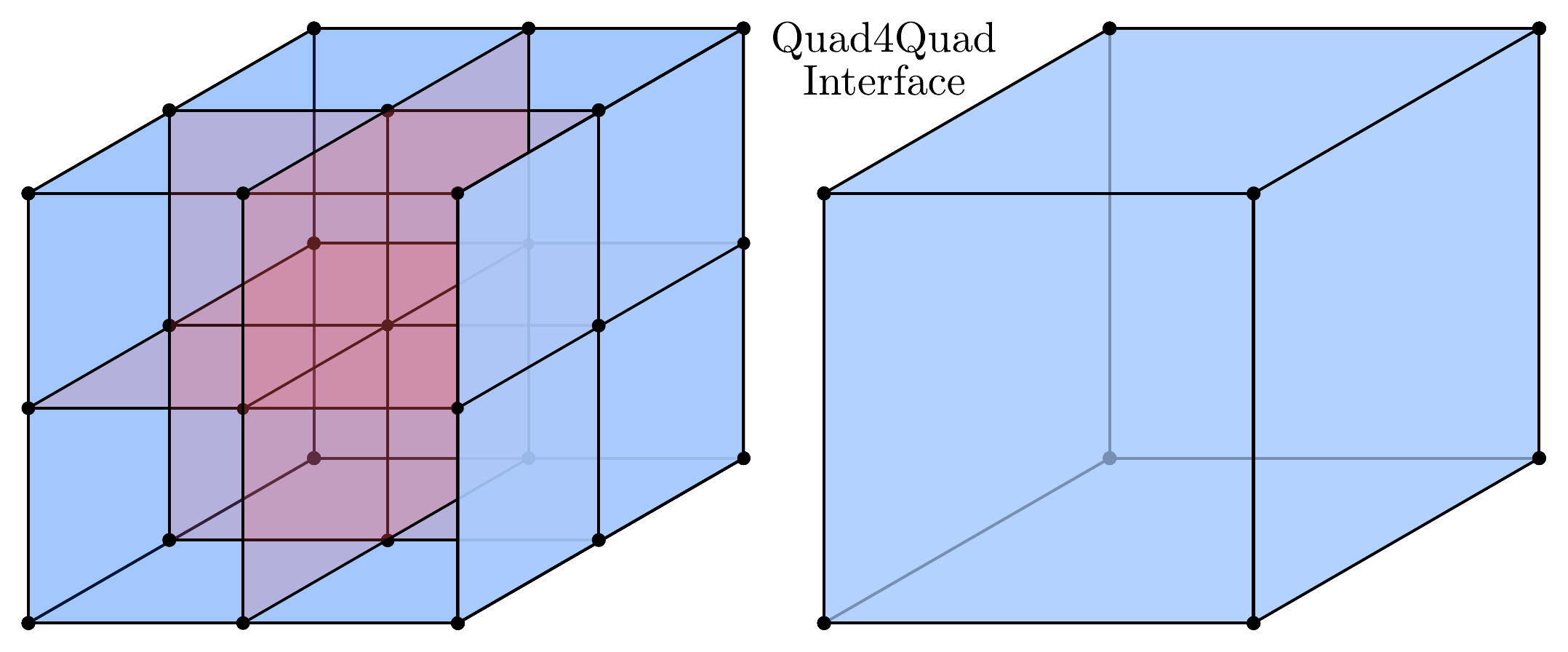}%
  \caption{Hanging nodes distribution in a Quad2Quad interface (left) and Quad4Quad interface (right).}%
  \label{fig:mesh:non-conforming-interfaces}%
\end{figure}
\begin{itemize}
  \item {\textbf{Refinement of hexahedral elements}. All flagged cells are split into $4$ or $8$ new children cells (for anisotropic or isotropic mesh refinement, respectively). Hanging nodes can appear, and the relation between the face nodes of neighboring elements with different levels of refinement has to be computed and exported in such a way that FSDM can handle them. For the hexahedral meshes used within the preprocessing tool and when hanging nodes are present, the Quad2Quad and Quad4Quad interfaces are computed (see \cref{fig:mesh:non-conforming-interfaces}). FSDM treats the hanging nodes in the following way: With help of pseudo element types, FSDM stores \enquote{hanging} connections. These are ignored by the solver, have no volume and no solution values, but at the same time they are used by the face extractor to create the face based grid. After that, FSDM completes the hanging node grid to a kind of (pseudo) conformity, which enable the adaptation to work on hanging elements in the same way (only with other element types) as on conforming elements.}
  \item {\textbf{Storing the final mesh in arrays}. In these arrays are stored the nodes, elements, conforming faces and non-conforming faces.}
  \item {Exporting the mesh arrays to HDF5 format, following the FSDM guidelines for the HDF5 format.}
\end{itemize}
\subsection{Postprocessing}\label{sec:postprocessing}
The postprocessing for the IBVP methodology is based on the computation of pressure coefficient, skin friction coefficient, lift coefficient and drag coefficient on the immersed body surface. Two different approaches are considered: integration over the STL and over a modified immersed body tessellation integration.
\subsubsection*{Integration over the Original Immersed Body Tessellation}
This approach takes advantage of the information within the STL file to perform the integration. First, for each triangle of the tessellation, the fluid variables are interpolated at the barycenter, at the vertices, and at the middle point of each edge of the triangle. The interpolation can be performed using different approaches: Radial Basis Functions (RBF), Inverse Distance, or Linear Interpolation. In general, a simple linear interpolation has shown good behavior and robustness.\par
Then the viscous stress tensor (including the pressure) $\tilde{\tensorsym{\tau}}=\tensorsym{\tau}-p\tensorsym{I}$ is computed at those points. Finally, the integration over each triangle is performed as follows:
\begin{equation}
  \vectorsym{F}_{t} = \frac{S}{60}\left(%
    27\left[\tilde{\tensorsym{\tau}}\cdot\hat{\vectorsym{n}}_{t}\right]_{\mathrm{bary}} +
    3\sum_{i=1}^{3}\left[\tilde{\tensorsym{\tau}}\cdot\hat{\vectorsym{n}}_{t}\right]_{\mathrm{vert},i} +
    8\sum_{i=1}^{3}\left[\tilde{\tensorsym{\tau}}\cdot\hat{\vectorsym{n}}_{t}\right]_{\mathrm{mid},i}\right).
\end{equation}
In this case, the integrated force $F_{t}$ on the triangle $t$ of area $S$ is computed as a weighted average of the stress tensor (including the pressure), $\tilde{\tensorsym{\tau}}$, projected over the normal vector of the local triangle. The value of $\left[\tilde{\tensorsym{\tau}}\cdot\hat{\vectorsym{n}}_{t}\right]_{\mathrm{bary}}$, $\left[\tilde{\tensorsym{\tau}}\cdot\hat{\vectorsym{n}}_{t}\right]_{\mathrm{vert},i}$, and $\left[\tilde{\tensorsym{\tau}}\cdot\hat{\vectorsym{n}}_{t}\right]_{\mathrm{mid},i}$ correspond to the projected force at the barycenter, at the vertex $i$ and at the middle point of the side $i$ respectively.
\subsubsection*{Integration over a Modified Immersed Body Tessellation}
This approach provides an alternative to compute the lift and drag when the STL is too coarse (and the integration over the STL shows a bad quality) or too fine (and the computational cost has to be reduced). In these cases, a new set of points on the surface of the immersed body are defined to perform the integration.\par
First, a region around the body is defined and every integration point of the mesh within that region is projected over the body's surface. Then, an algorithm is used to compute the weights for each point. Once the points on the surface (and their weights) are known, the fluid variables are interpolated on those points. Finally, the stress tensor (including the pressure) is computed following the same equations shown above and the projected force is integrated as follows:
\begin{equation}
  F = \sum_{i=0}^{\mathrm{nPoints}}\tilde{{\tau}}_{i}\cdot\hat{\vectorsym{n}}_{i}\omega_{i}.
\end{equation}
In this case, the global force is computed as the dot product between the projected force at each point, $\tilde{\tensorsym{\tau}}$, and the weights, $\omega$.
%
%
\section{Numerical Computations}\label{sec:numerical-computations}
To prove the validity and performance of this approach,  we  present in this section several numerical simulations of different fluid flow problems of increasing difficulty. These tests are the subsonic flow around the NACA0012 airfoil at different angles of attack, the transonic flow around the RAE2822 airfoil, the subsonic flow around the MDA30P30N multi-element airfoil, and the more challenging NASA High-Lift Common Research Model (CRM-HL) aircraft. Experimental results of most of those problems are reported in the literature and thus will serve as validation of our methodology.  For all simulations, the RANS equations were used for modeling the fluid flow, and the numerical methods employed by CODA were, for the spatial discretization, the second order finite volume method coupled with the immersed boundary volume penalization. A linearized implicit Euler scheme is employed for time discretization. All simulations were executed using $\num{600}$ cores in the Magerit Supercomputer at Supercomputing and Visualization Center of Madrid (CeSViMa) and $\num{3840}$ cores in the Marenostrum4 Supercomputer at the Barcelona Supercomputing Center.
%
\subsection{Flow around the NACA0012 airfoil}
\begin{figure}[h]
  \centering
  \includegraphics[width=0.49\linewidth]{\figurespath/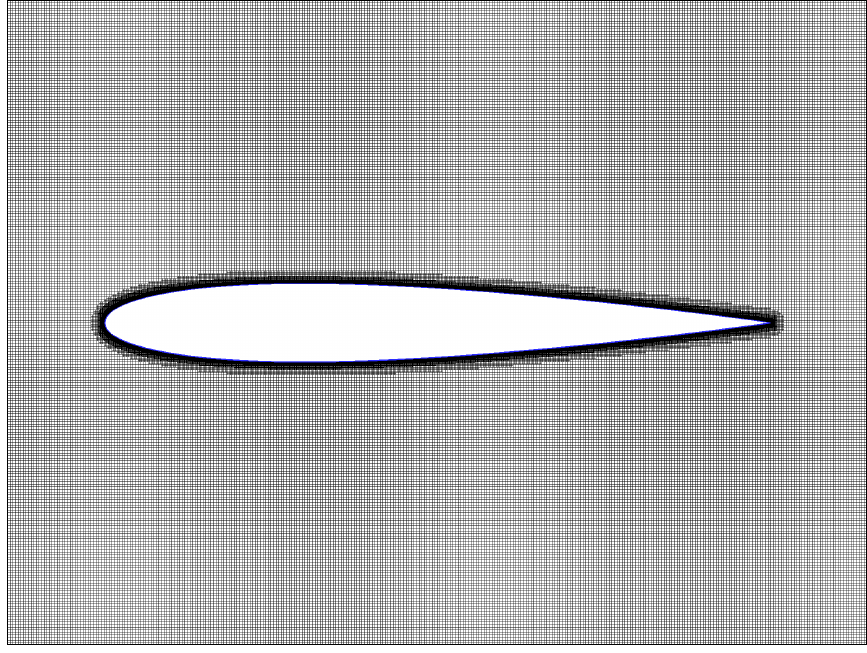}%
  \hfill%
  \includegraphics[width=0.49\linewidth]{\figurespath/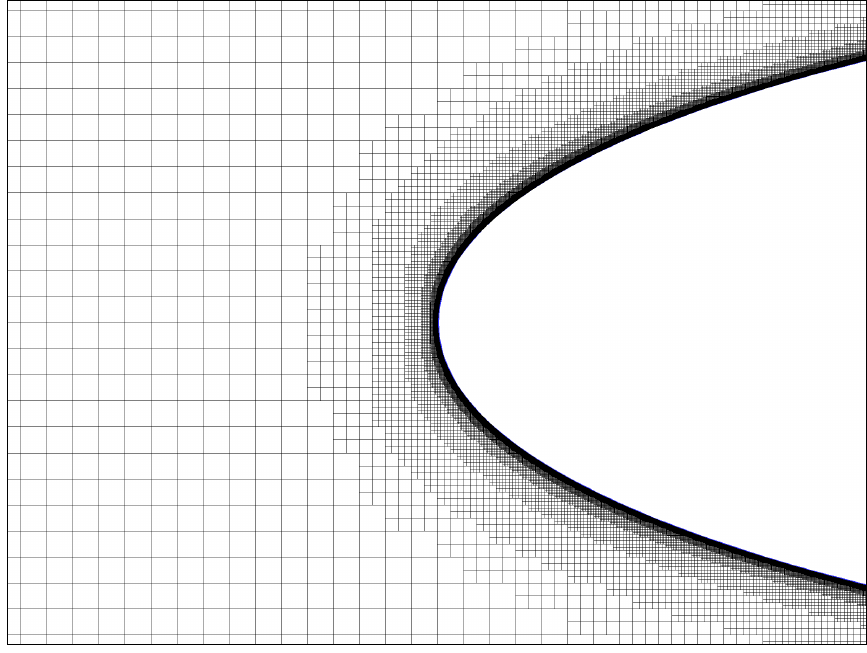}%
  \\[\medskipamount]
  \caption{Mesh refinement around the NACA0012 airfoil. The mesh has been generated with maximum level of refinement $l_{\mathrm{max}}=16$ and with the blanking activated. Close-up view of the mesh at $50x$ zoom, depicting the full airfoil (left) and close-up view of the mesh at $500x$ zoom, depicting the leading edge (right).}%
  \label{fig:numerical-computations:naca0012:mesh1}%
\end{figure}

\begin{figure}[h]
  \centering
  \includegraphics[width=0.49\linewidth]{\figurespath/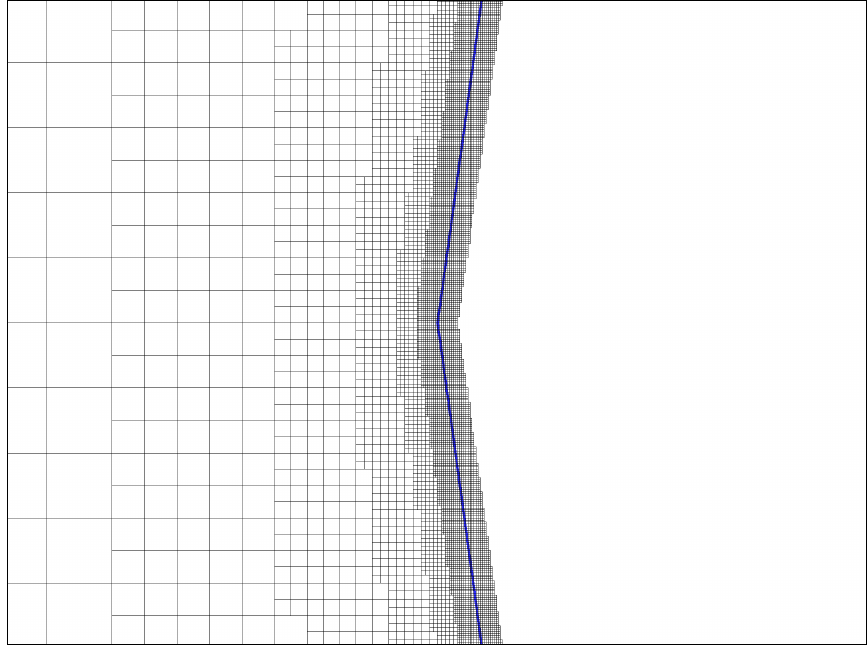}%
  \hfill%
  \includegraphics[width=0.49\linewidth]{\figurespath/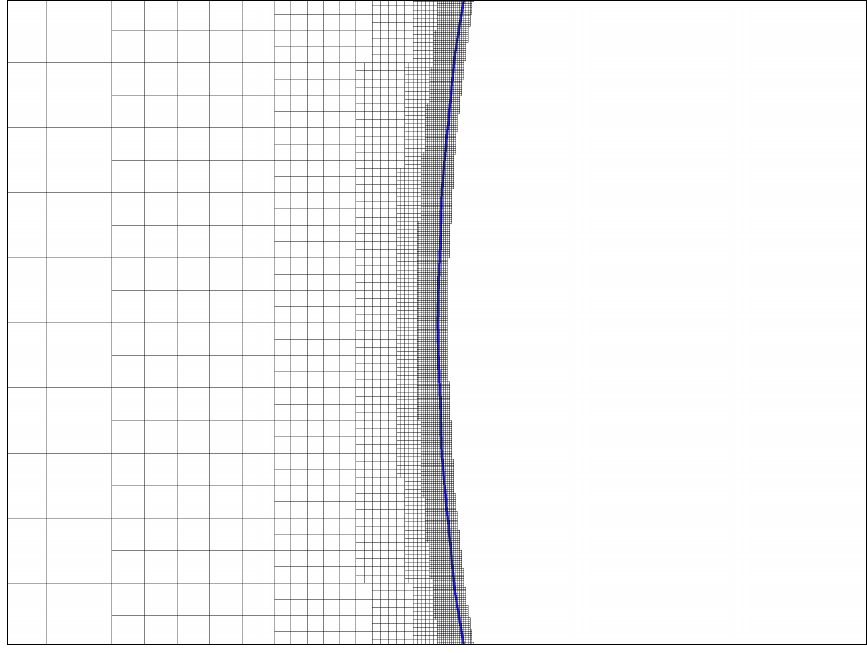}%
  \\[\medskipamount]
  \caption{Mesh refinement around the NACA0012 airfoil. The meshes have been generated with maximum level of refinement $l_{\mathrm{max}}=16$ and with the blanking activated. Close-up view of the meshes at $10000x$ zoom, depicting the leading edge and using a low-resolution STL geometry (left) and a high-resolution STL geometry (right).}%
  \label{fig:numerical-computations:naca0012:mesh2}%
\end{figure}

The computational domain for the fluid flow around the NACA0012 airfoil is the box with dimensions $[-20,+20]\times[-20,+20]\times[0,+1]$. The domain span is sufficient to avoid domain confinement effects and wave reflections from the domain boundaries, which could lead to a significant error when computing the lift and drag forces around the airfoil. The extent of the computational domain corresponds to $40c$, where $c$ is the chord length and is equal to $1$. The domain is initially discretized with $n_{x}n_{y}n_{z}=1600$ hexahedral elements, where $n_{x}=40$, $n_{y}=40$, and $n_{z}=1$ are the number of elements in the direction $x$, $y$, and $z$, respectively. This initial mesh is further refined around the immersed geometry and also in the wake region. The spacing of the grid obeys $h=2^{-l}$, where $l$ is the level of refinement. Sufficient grid resolution around the airfoil is crucial to obtain accurate results for the lift and drag coefficients. The mesh around the immersed geometry is refined up to the level of refinement $l=16$, we get a mesh with around $\num{4.99E6}$ elements. The wake region is resolved with a refined subregion with refinement level $l=8$. This region is approximately $20$ airfoil chords downstream of the leading edge of the airfoil. The characteristics of the meshes used for the mesh convergence study are shown in \cref{tab:numerical-computations:naca0012:mesh}; ranging from a coarse mesh to a very fine mesh. The refinement level $l_{\mathrm{max}}$ listed in the table is the highest level of refinement around the immersed body.
\begin{figure}[h]
  \centering
  \includegraphics[width=0.49\linewidth]{\figurespath/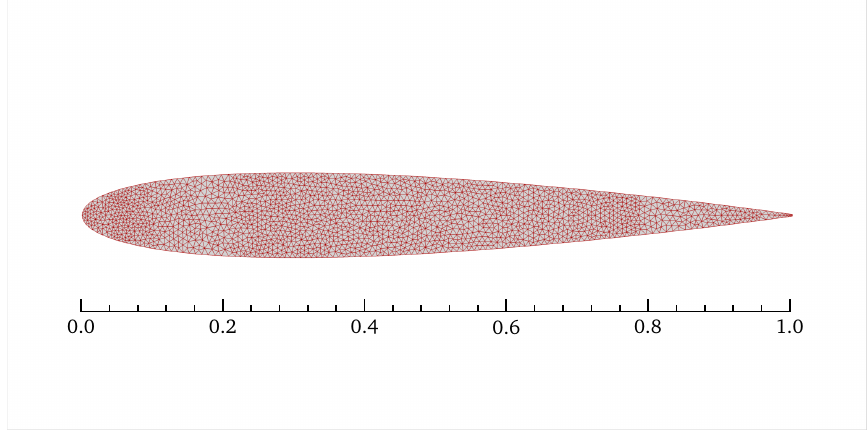}%
  \hfill%
  \includegraphics[width=0.49\linewidth]{\figurespath/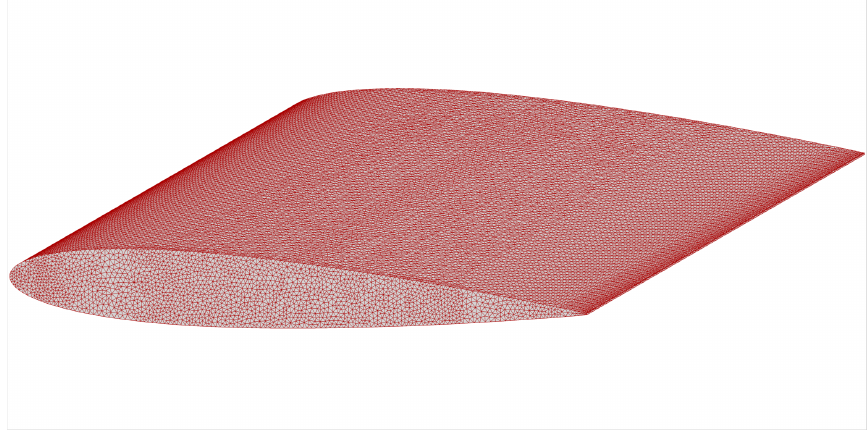}%
  \\[\medskipamount]
  \caption{Geometry used for the numerical simulations of the NACA0012 airfoil.}%
  \label{fig:numerical-computations:naca0012:geometry}%
\end{figure}
\begin{table}[h]
  \setlength{\fboxsep}{0.00pt}%
  \renewcommand{\arraystretch}{1.2}%
  \newcolumntype{A}{X}%
  \newcolumntype{B}{p{0.2\textwidth}}%
  \caption{Characteristics of the meshes for the flow around the NACA0012 airfoil.}%
  \label{tab:numerical-computations:naca0012:mesh}%
  \begin{tabularx}{\textwidth}{AAAAX}
    \arrayrulecolor{black}\hline
    {Name} &
    {$l_{\mathrm{max}}$} &
    {$h_{\mathrm{wake}}$} &
    {$h_{\mathrm{body}}$} &
    {$\mathit{nElems}$} \\
    \arrayrulecolor{black}\hline
    Coarse       & $10$ & $\num{3.91E-03}$ & $\num{9.77E-04}$ & $\num{6.53E+05}$ \\
    Medium       & $12$ & $\num{3.91E-03}$ & $\num{2.44E-04}$ & $\num{8.59E+05}$ \\
    Fine         & $14$ & $\num{3.91E-03}$ & $\num{6.10E-05}$ & $\num{1.68E+06}$ \\
    Very Fine    & $16$ & $\num{3.91E-03}$ & $\num{1.53E-05}$ & $\num{4.99E+06}$ \\
    \arrayrulecolor{black}\hline
  \end{tabularx}
\end{table}
The immersed geometry is located in the center of the computational domain, with its cross-section in the plane $z$, and this corresponds to the NACA0012 profile with chord length $c=1$. The triangulation of the surface geometry is made up of $\num{65468}$ triangles for the low resolution STL and $\num{4720046}$ triangles for the high resolution STL. The geometry is shown in \cref{fig:numerical-computations:naca0012:geometry}, and in \cref{fig:numerical-computations:naca0012:mesh1} the mesh around the immersed geometry is depicted. In \cref{fig:numerical-computations:naca0012:mesh2} are shown, at $10000x$ zoom, the meshes around the NACA0012 airfoil with maximum level of refinement $l_{\mathrm{max}}=16$ and with the blanking activated; the leading edge is depicted for a low-resolution STL geometry (left) and a high-resolution STL geometry (right).\par
\subsubsection{Subsonic flow at $M_{\infty}=\num{0.15}$ and $\alpha=\SI{0}{\degree}$ and $\alpha=\SI{10}{\degree}$}
The first test corresponds to the subsonic flow around an NACA0012 airfoil. This problem has become a classic test case for RANS solvers due to the simple geometry and the large amount of available numerical and experimental data \cite{ladson1988a,ladson1988b,ladson1988c}. It is used primarily for the analysis of turbulence models, testing their convergence properties and their effect on the accuracy. We perform simulations for the angle of attack $\alpha=\SI{0}{\degree}$ and $\alpha=\SI{10}{\degree}$. The flow satisfies the following conditions: a gas with adiabatic index $\gamma=1.4$ and Reynolds number $\mathrm{Re}=\num{6E6}$ flows with a free-stream Mach number $M_{\infty}=\num{0.15}$. The non-dimensional density is set to $\rho=\num{1}$, and the non-dimensional pressure to $p=\num{1}$. On the boundaries of the computational domain the following boundary conditions were set: the left face is set to inflow, the right face to outflow, the front and back faces were set to periodic, and the top and bottom to far field. The inflow pressure assumes the value of the stagnation pressure and the outflow pressure to the static pressure.\par
\begin{figure}[h]
  \centering
  \includegraphics[width=0.49\linewidth]{\figurespath/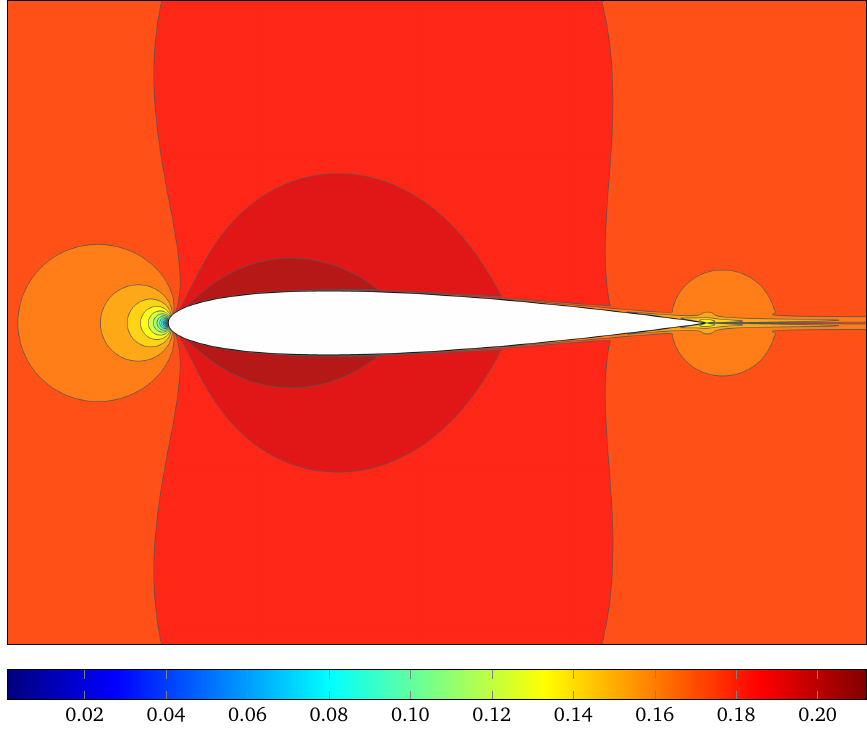}%
  \hfill%
  \includegraphics[width=0.49\linewidth]{\figurespath/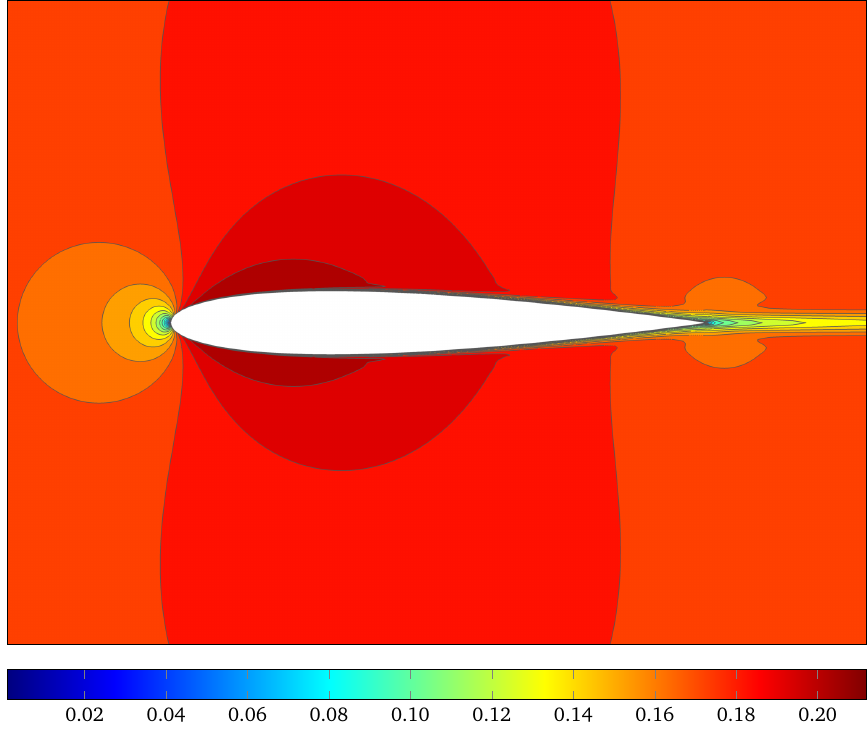}%
  \\[\medskipamount]
  \caption{Subsonic flow around the NACA0012 airfoil. Contour plots of the velocity magnitude for the simulations with angle of attack $\alpha=\SI{0}{\degree}$. Computations were performed using the CODA solver based on a body-fitted mesh (left) and an IBM Cartesian mesh with maximum level of refinement $l_{\mathrm{max}}=16$ (right). In both computations, the RANS equations were solved with the wall model deactivated.}%
  \label{fig:numerical-computations:naca0012:subsonic:aoa-0:velocity1}%
\end{figure}
\begin{figure}[h]
  \centering
  \includegraphics[width=0.49\linewidth]{\figurespath/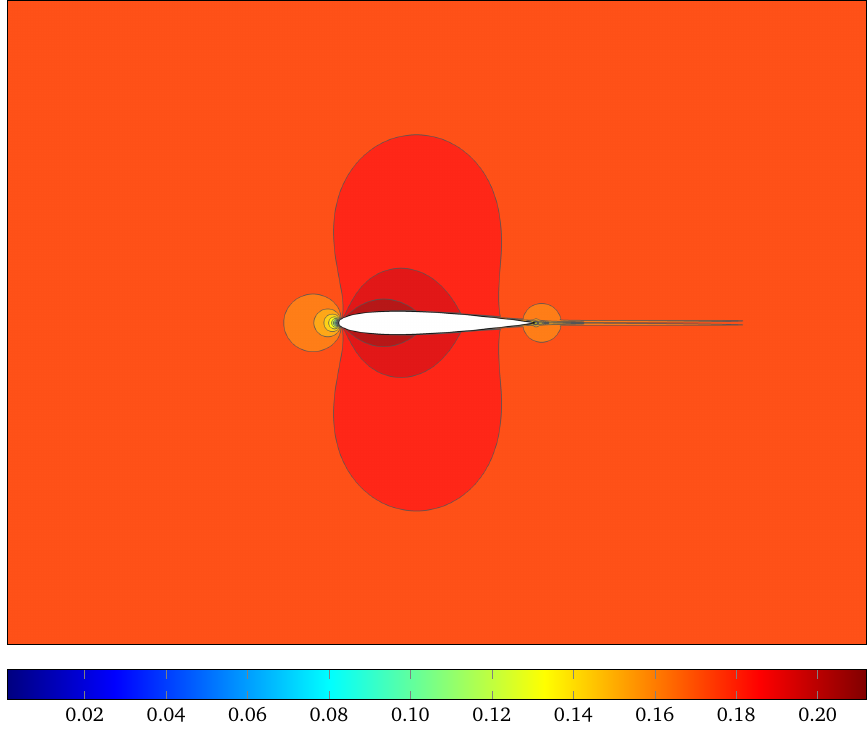}%
  \hfil%
  \includegraphics[width=0.49\linewidth]{\figurespath/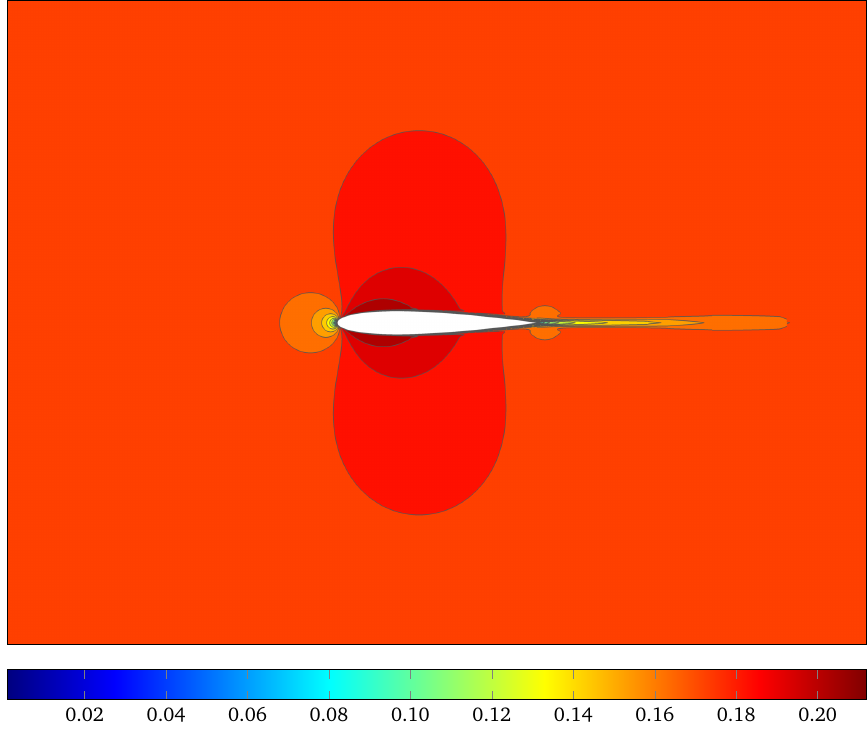}%
  \\[\medskipamount]
  \caption{Subsonic flow around the NACA0012 airfoil. Contour plots of the velocity magnitude for the simulations with angle of attack $\alpha=\SI{0}{\degree}$. Computations were performed using the CODA solver based on a body-fitted mesh (left) and an IBM Cartesian mesh with maximum level of refinement $l_{\mathrm{max}}=16$ (right). In both computations, the RANS equations were solved with the wall model deactivated.}%
  \label{fig:numerical-computations:naca0012:subsonic:aoa-0:velocity2}%
\end{figure}
\begin{figure}[h]
  \centering
  \includegraphics[width=0.49\linewidth]{\figurespath/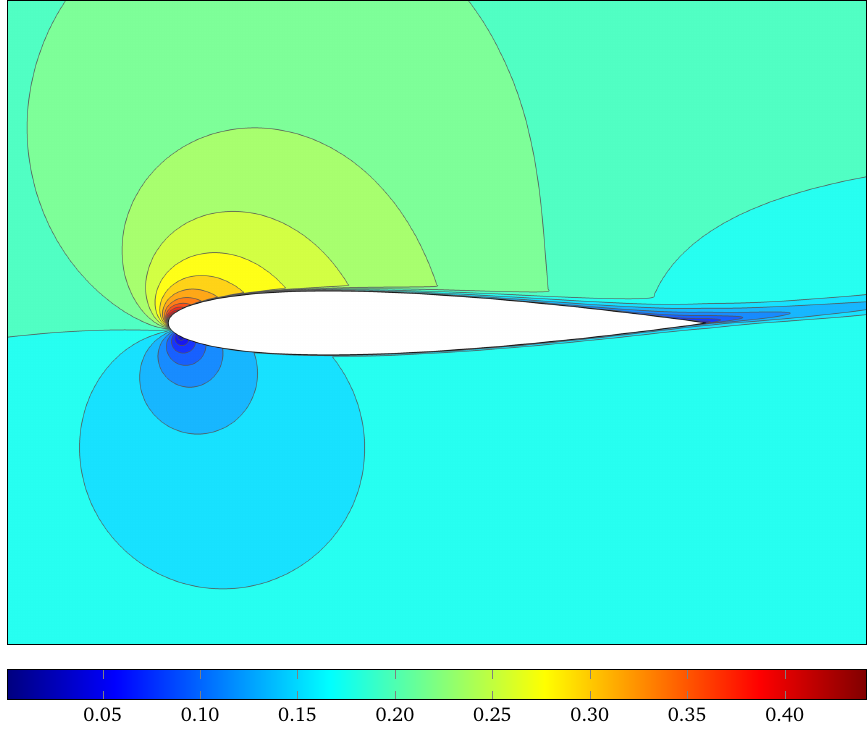}%
  \hfil%
  \includegraphics[width=0.49\linewidth]{\figurespath/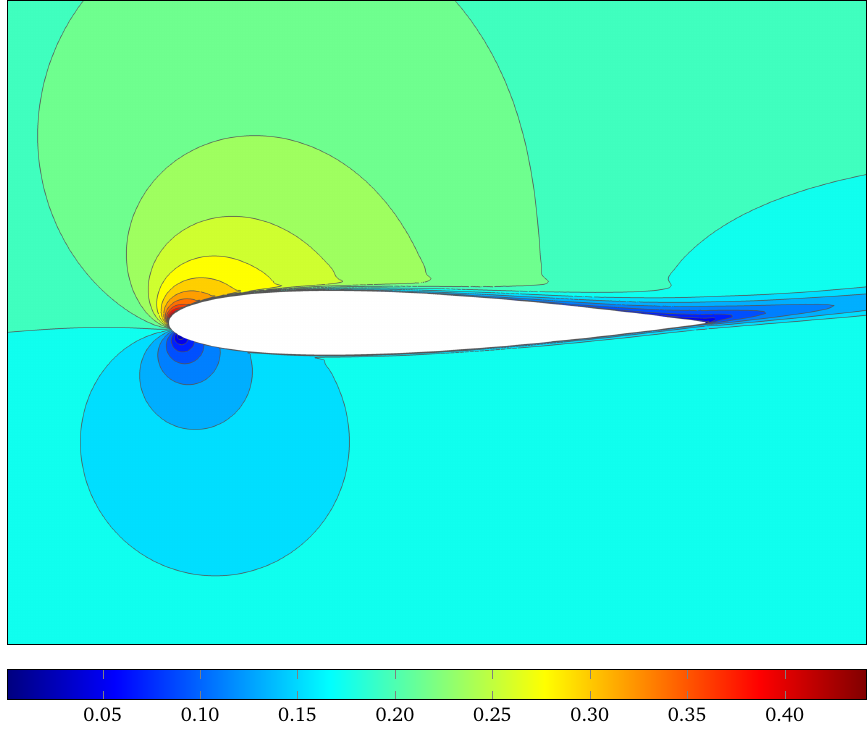}%
  \\[\medskipamount]
  \caption{Subsonic flow around the NACA0012 airfoil. Contour plots of the velocity magnitude for the simulations with angle of attack $\alpha=\SI{10}{\degree}$. Computations were performed using the CODA solver based on a body-fitted mesh (left) and an IBM Cartesian mesh with maximum level of refinement $l_{\mathrm{max}}=16$ (right). In both computations, the RANS equations were solved with the wall model deactivated.}%
  \label{fig:numerical-computations:naca0012:subsonic:aoa-10:velocity}%
\end{figure}
In the \cref{fig:numerical-computations:naca0012:subsonic:aoa-0:velocity1} and \cref{fig:numerical-computations:naca0012:subsonic:aoa-0:velocity2} are depicted the contour plots of the velocity magnitude for the flow past the NACA0012 profile at angle of attack $\alpha=\SI{0}{\degree}$. The mesh employed in the simulation is the very fine mesh with element's size around the body $h_{\mathrm{body}}=\num{1.53E-05}$, element's size in the wake region $h_{\mathrm{wake}}=\num{3.91e-3}$. The Spallart--Allmaras model has been activated. We observe the flow field is symmetric as the fluid past the airfoil. As the fluid flows over the blunt trailing edge, a minor amount of flow separation occurs, but it reattaches shortly after it passes the trailing edge and recovers the mean velocity in the distant far field.\par
In the \cref{fig:numerical-computations:naca0012:subsonic:aoa-10:velocity} are depicted the contour plots of the velocity magnitude for the flow past the NACA0012 profile at angle of attack $\alpha=\SI{10}{\degree}$. The mesh employed is the same as for angle of attack $\alpha=\SI{0}{\degree}$. The Spallart--Allmaras model has been activated. We observe that the stagnation point is shifted downward (relative to the airfoil), reflective of the shifted angle of attack. The fluid velocity above the airfoil is clearly different than that below the airfoil, showing a low-velocity region toward the upper side of the trailing edge, which means flow separation.\par
\begin{figure}[h]
  \centering
  \includegraphics[width=0.49\linewidth]{\figurespath/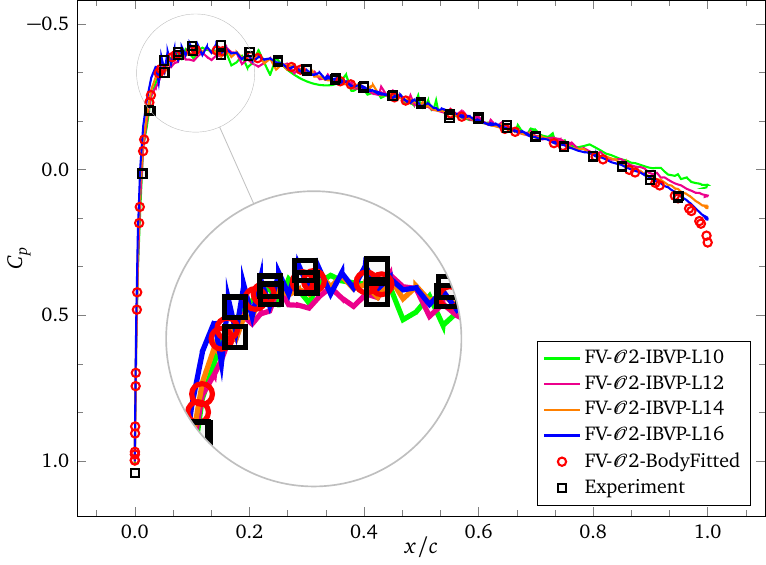}%
  \hfil%
  \includegraphics[width=0.49\linewidth]{\figurespath/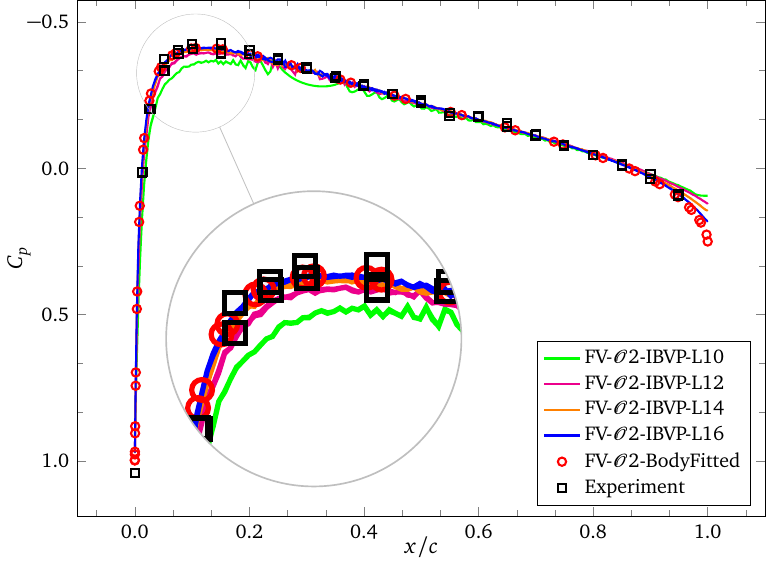}%
  \\[\medskipamount]
  \caption{Subsonic flow around the NACA0012 airfoil. Grid convergence study in terms of the pressure coefficient $C_{p}$ for the subsonic flow over the NACA0012 airfoil surface at angle of attack $\alpha=\SI{0}{\degree}$ for the cases with low-resolution STL geometry (left) and high-resolution STL geometry (right). In both computations, the RANS equations were solved with the wall model deactivated.}%
  \label{fig:numerical-computations:naca0012:subsonic:aoa-0:cp1}%
\end{figure}
\begin{figure}[h]
  \centering
  \includegraphics[width=0.49\linewidth]{\figurespath/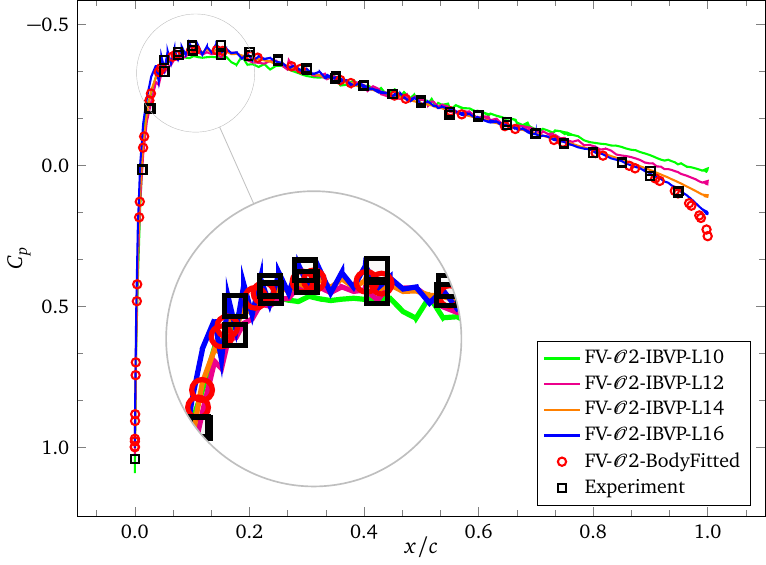}%
  \hfil%
  \includegraphics[width=0.49\linewidth]{\figurespath/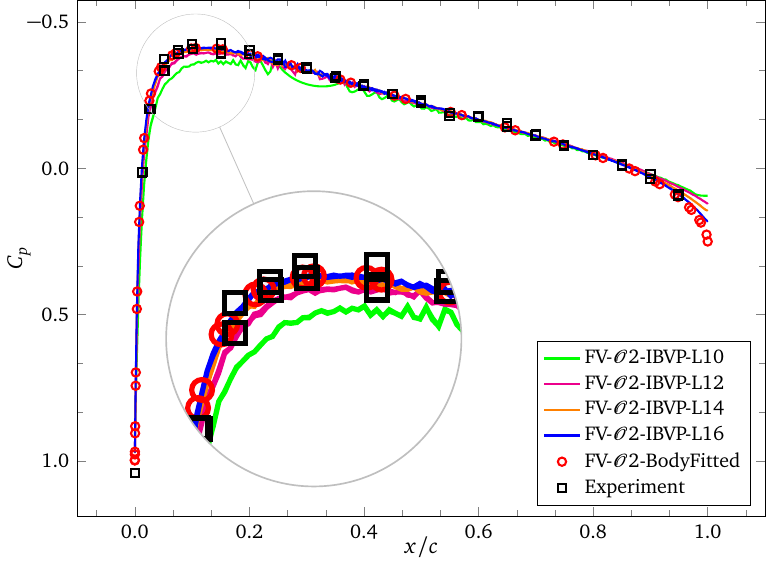}%
  \\[\medskipamount]
  \caption{Subsonic flow around the NACA0012 airfoil. Grid convergence study in terms of the pressure coefficient $C_{p}$ for the subsonic flow over the NACA0012 airfoil surface at angle of attack $\alpha=\SI{0}{\degree}$ for the cases with low-resolution STL geometry (left) and high-resolution STL geometry (right). In both computations, the RANS equations were solved with the wall model activated.}%
  \label{fig:numerical-computations:naca0012:subsonic:aoa-0:cp2}%
\end{figure}
\begin{figure}[h]
  \centering
  \includegraphics[width=0.49\linewidth]{\figurespath/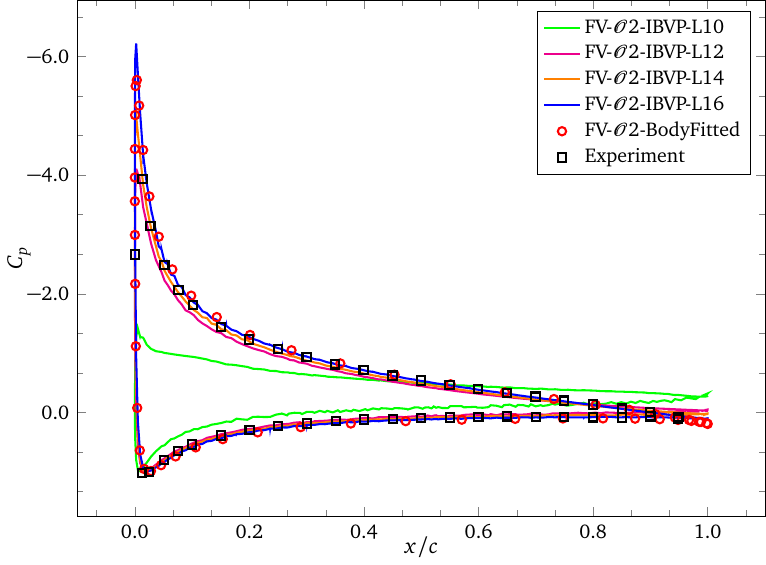}%
  \hfil%
  \includegraphics[width=0.49\linewidth]{\figurespath/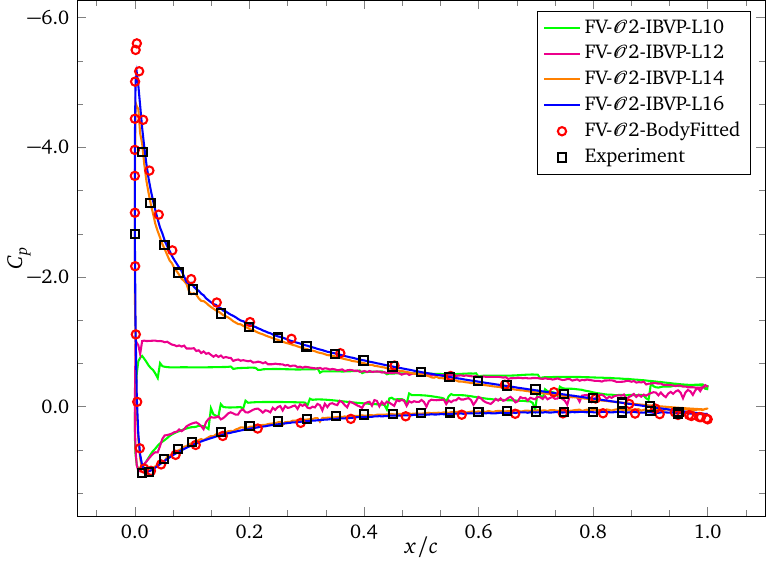}%
  \\[\medskipamount]
  \caption{Subsonic flow around the NACA0012 airfoil. Grid convergence study in terms of the pressure coefficient $C_{p}$ for the subsonic flow over the NACA0012 airfoil surface at angle of attack $\alpha=\SI{10}{\degree}$ for the cases with low-resolution STL geometry (left) and high-resolution STL geometry (right). In both computations, the RANS equations were solved with the wall model deactivated.}%
  \label{fig:numerical-computations:naca0012:subsonic:aoa-10:cp}%
\end{figure}
Now we analyze the pressure coefficient plots. The \cref{fig:numerical-computations:naca0012:subsonic:aoa-0:cp1,fig:numerical-computations:naca0012:subsonic:aoa-0:cp2} show the pressure coefficient $C_{p}$ on the airfoil surface for the simulation case with an angle of attack $\alpha=\SI{0}{\degree}$. All meshes in \cref{tab:numerical-computations:naca0012:mesh} have been used and the cases where the STL geometry has a low resolution and a high resolution. Computations with a deactivated and activated wall model have been considered. The experimental data are also plotted and were obtained from \cite{ladson1988a}. For an angle of attack $\alpha=\SI{0}{\degree}$, and for all considered meshes, for computations with a low-resolution STL geometry, the pressure coefficient curve is very close to the experimental data, but some unphysical oscillations are clearly perceptible in all simulations, both with the wall model switched off and switched on. These oscillations tend to diminish as the mesh becomes finer. For simulations with high-resolution geometry, the oscillations are present only for the coarser mesh. The STL geometry resolution has an important impact on the accuracy of the computations along with the mesh refinement. We can observe that the computations with the wall model deactivated and activated, and for high-resolution STL geometries, show very similar $C_{p}$ curves, suggesting that for good enough resololutions (in terms of volume mesh and surface tessellation of the STL), wall functions are not necessary.\par
In \cref{fig:numerical-computations:naca0012:subsonic:aoa-10:cp} is represented by the pressure coefficient $C_{p}$ on the airfoil surface for an angle of attack $\alpha=\SI{10}{\degree}$ and for all meshes reported in \cref{tab:numerical-computations:naca0012:mesh}. Only the cases with deactivated wall model have been taken into account. The pressure coefficient curves for the coarse and medium meshes show a large discrepancy with respect to the experimental data, especially on the upper surface of the leading edge. Only the fine and very fine meshes are in good agreement with the experiment. The computations with high-resolution STL geometries show an oscillatory behavior for the coarse and medium meshes. For simulations with low-resolution meshes, at leading edge, the $C_{p}$ seems to diverge, but this divergence is not present for the high-resolution STL geometry computations. Although $C_{p}$ curves for computations with wall model activated are not shown, the diverging behavior is also present in the cases with fine and very fine meshes. We are convinced that geometry and mesh resolutions play an important role in simulating turbulent flows accurately with IBVP methods.\par
\begin{figure}[h]
  \centering
  \includegraphics[width=0.49\linewidth]{\figurespath/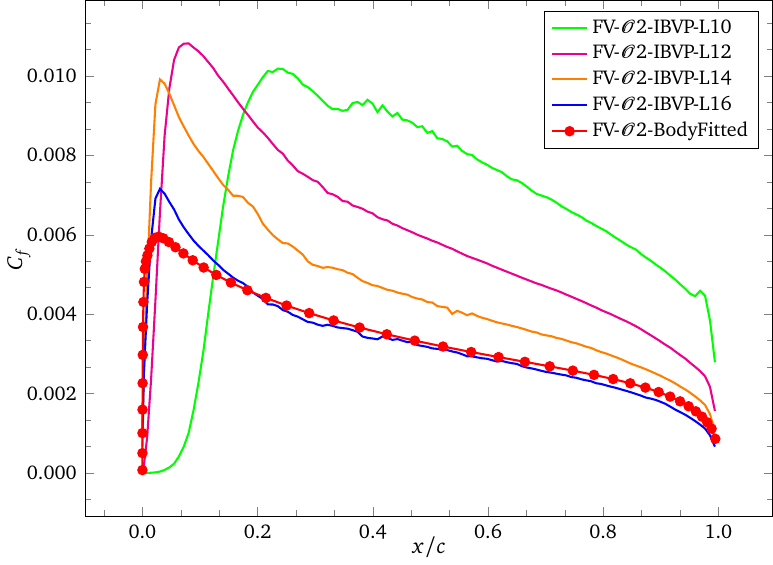}%
  \hfil%
  \includegraphics[width=0.49\linewidth]{\figurespath/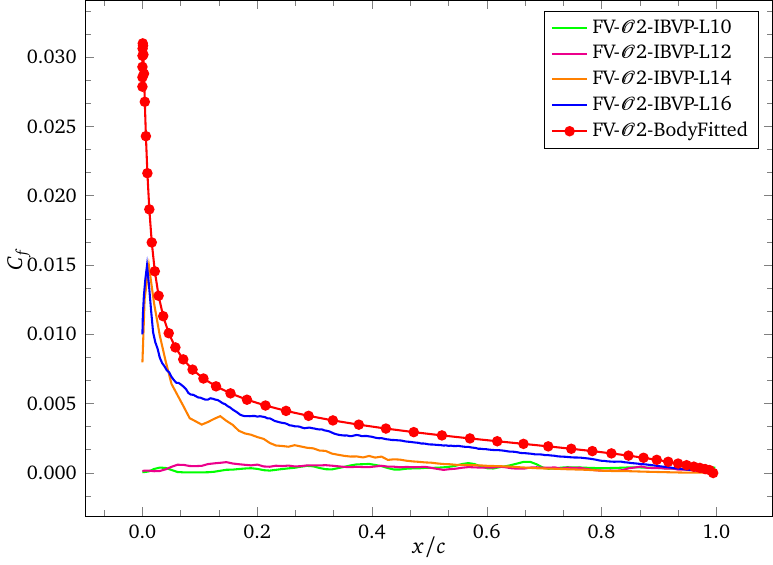}%
  \\[\medskipamount]
  \caption{Subsonic flow around the NACA0012 airfoil. Grid convergence study in terms of the skin friction coefficient $C_{f}$ for the subsonic flow over the NACA0012 airfoil surface at angles of attack $\alpha=\SI{0}{\degree}$ (left) and $\alpha=\SI{10}{\degree}$ (right) and employing a high-resolution STL geometry. In both computations, the RANS equations were solved with the wall model deactivated.}%
  \label{fig:numerical-computations:naca0012:subsonic:aoa-00-10:cf}%
\end{figure}
In \cref{fig:numerical-computations:naca0012:subsonic:aoa-00-10:cf} we show the skin friction coefficient curves at $z=0.5$. We observe that the surface skin friction coefficient is not very well predicted as the pressure coefficient. However, as the resolution of the mesh increases, $C_{f}$ is approaching to the reference values taken from the body-fitted simulations. This discrepancy is typically attributed to the way the gradients are interpolated. An appropriate way to reconstruct $C_{f}$ is still an open question. Several strategies have been proposed in \cite{goza2016a} and are worth investigating in future works.\par
The lift and drag values are shown in \cref{tab:numerical-computations:naca0012:subsonic:aoa-00-10:lift-drag} for the angle of attack $\alpha=\SI{0}{\degree}$ and $\alpha=\SI{10}{\degree}$. For the simulation with angle of attack $\alpha=\SI{10}{\degree}$, the lift coefficient $C_{L}$ increases monotonically as the grid is refined, getting closer to the experimental value, while the drag coefficient $C_{D}$ approaches its numerical value to the experimental one. This behavior occurs for simulations with the wall model deactivated and activated. For the simulation with angle of attack $\alpha=\SI{0}{\degree}$, we do not observe monotonicity in the evolution of the lift coefficient $C_{L}$ and the drag coefficient $C_{D}$ as the mesh is refined, however, as the mesh is refined, the $C_{L}$ and $C_{D}$ approach to the body-fitted and experimental values.
\begin{table}[h]
  \setlength{\fboxsep}{0.00pt}%
  \renewcommand{\arraystretch}{1.2}%
  \newcolumntype{A}{C}%
  \newcolumntype{B}{p{0.2\textwidth}}%
  \caption{Subsonic flow around the NACA0012 airfoil. Lift and drag coefficients for angle of attack $\alpha=\SI{0}{\degree}$ and $\alpha=\SI{10}{\degree}$.}%
  \label{tab:numerical-computations:naca0012:subsonic:aoa-00-10:lift-drag}%
  \begin{tabularx}{\textwidth}{XAAAA}
    \arrayrulecolor{black}\hline
    \multicolumn{1}{l}{} &
    \multicolumn{2}{c}{$\alpha=\SI{0}{\degree}$} &
    \multicolumn{2}{c}{$\alpha=\SI{10}{\degree}$} \\
    \cmidrule(r){2-3}
    \cmidrule(l){4-5}
    Mesh &
    $C_{L}$ &
    $C_{D}$ &
    $C_{L}$ &
    $C_{D}$ \\
    \arrayrulecolor{black}\hline
    Coarse       & $\num{4.381E-06}$ & $\num{1.630E-02}$ & $\num{2.031E-01}$ & $\num{9.448E-02}$ \\
    Medium       & $\num{2.472E-06}$ & $\num{1.213E-02}$ & $\num{2.511E-01}$ & $\num{1.000E-01}$ \\
    Fine         & $\num{6.884E-06}$ & $\num{8.672E-03}$ & $\num{4.627E-01}$ & $\num{9.159E-02}$ \\
    Very Fine    & $\num{2.078E-05}$ & $\num{6.147E-03}$ & $\num{4.921E-01}$ & $\num{9.323E-02}$ \\
    Body-Fitted  & $\num{9.430E-06}$ & $\num{8.330E-03}$ & $\num[print-zero-exponent=true]{1.090E+00}$ & $\num{1.230E-02}$ \\
    Experimental & $\num{7.240E-03}$ & $\num{8.090E-03}$ & $\num[print-zero-exponent=true]{1.057E+00}$ & $\num{1.190E-02}$ \\
    \arrayrulecolor{black}\hline
  \end{tabularx}
\end{table}
\subsubsection{Influence of mesh blanking on simulations performance}
\begin{figure}[h]
  \centering
  \includegraphics[width=0.49\linewidth]{\figurespath/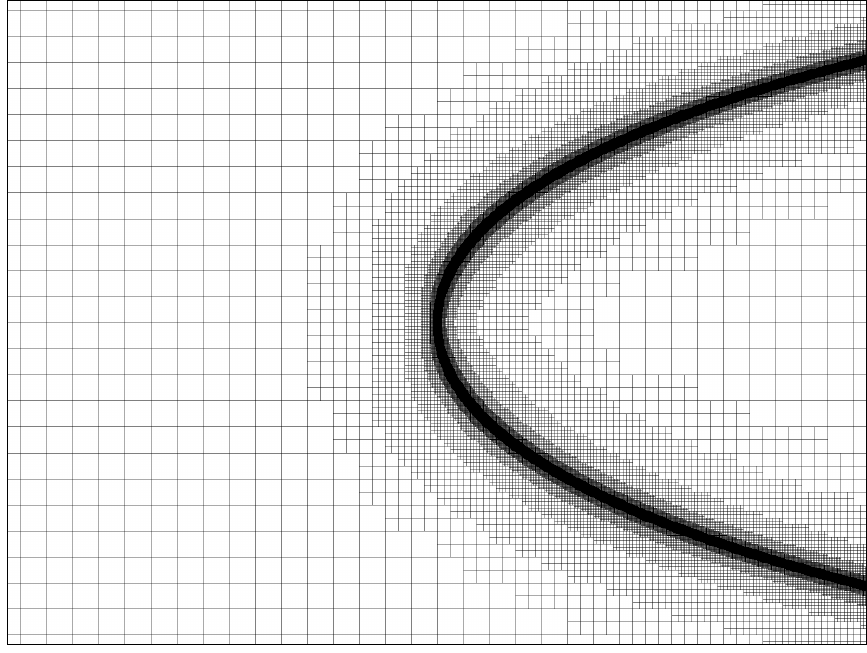}%
  \hfil%
  \includegraphics[width=0.49\linewidth]{\figurespath/NACA0012-BLANKING-ON-HR-leading-edge-z00500.pdf}%
  \\[\medskipamount]
  \includegraphics[width=0.49\linewidth]{\figurespath/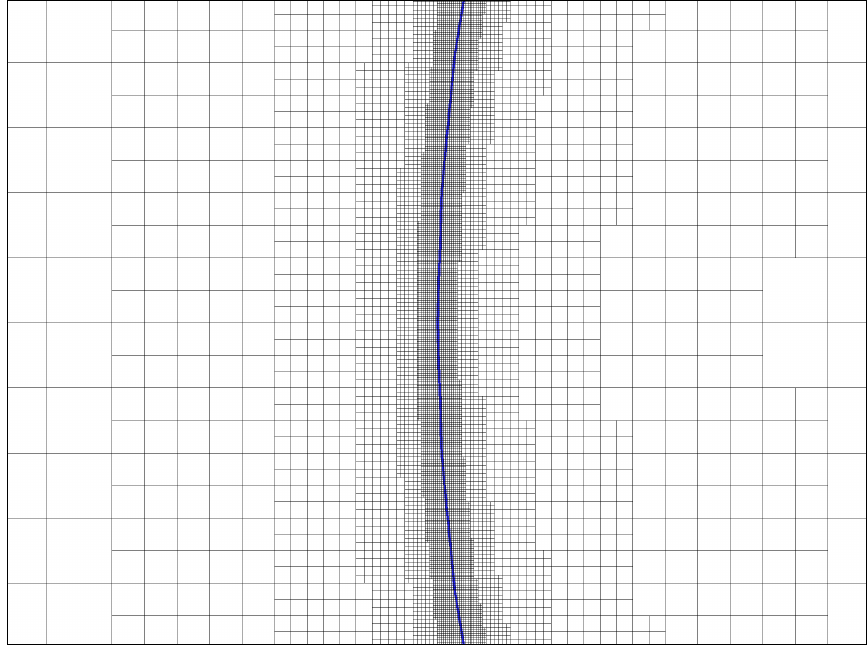}%
  \hfil%
  \includegraphics[width=0.49\linewidth]{\figurespath/NACA0012-BLANKING-ON-HR-leading-edge-z10000.pdf}%
  \\[\medskipamount]
  \caption{Mesh refinement around the NACA0012 airfoil without blanking of elements inside the geometry (left) and with blanking of elements inside the geometry (right). The blue lines represent the edges of the STL triangles. In both cases, a high-resolution STL geometry has been employed.}%
  \label{fig:numerical-computations:naca0012:mesh:blanking}%
\end{figure}

In the context of immersed boundary methods, the use of Cartesian meshes with refinement around the immersed geometry involves the generation of grids with a large number of elements. This represents a fundamental disadvantage with respect to CFD solvers based on body-fitted meshes. Mesh blanking in IBM Cartesian meshes is a procedure used to reduce the computation time when similar results are sought among CFD solvers based on IBM techniques and body-fitted meshes. In this work, we explore the mesh blanking to assess the performance of the CODA solver with IBM. In blanked meshes, elements within the immersed geometry are removed, leaving a thin layer of elements inside the body (see \cref{fig:numerical-computations:naca0012:mesh:blanking}). In \cref{tab:numerical-computations:naca0012:mesh:blanking} is shown the performance comparison between simulations of the NACA0012 subsonic flow at angle of attack $\alpha=\SI{0}{\degree}$ with immersed boundary methods and with meshes without blanking and with blanking. Computations done using blanked meshes have a superior efficiency compared to simulations using meshes without blanking. We can see that computations using blanked meshes require two orders of magnitude fewer linear iterations than computations with no blanked meshes in order to converge to a solution with similar residual in the density in computations performed with body-fitted meshes.
\begin{table}[h]
  \setlength{\fboxsep}{0.00pt}%
  \renewcommand{\arraystretch}{1.2}%
  \newcolumntype{A}{>{\raggedleft}p{1.20cm}}%
  \newcolumntype{B}{>{\raggedleft}X}%
  \caption{Subsonic flow around the NACA0012 airfoil. Influence of mesh blanking on simulations performance for angle of attack $\alpha=\SI{0}{\degree}$.}%
  \label{tab:numerical-computations:naca0012:mesh:blanking}%
  \begin{tabularx}{\textwidth}{XXAAXXAAX}
    \arrayrulecolor{black}\hline
    \multicolumn{1}{l}{} &
    \multicolumn{4}{c}{Blanking Deactivated} &
    \multicolumn{4}{c}{Blanking Activated} \\
    \cmidrule(r){2-5}
    \cmidrule(l){6-9}
    {Mesh} &
    {nElements} &
    {Time $\left[\SI{}{\hour}\right]$} &
    {nIter} &
    {Residual $\left[\rho\right]$} &
    {nElements} &
    {Time $\left[\SI{}{\hour}\right]$} &
    {nIter} &
    {Residual $\left[\rho\right]$} \\
    \arrayrulecolor{black}\hline
    Coarse      & $\num{6.53E+05}$ & $\num{6.68}$  & $\num{3000}$ & $\num{9.32E-06}$
                & $\num{7.19E+05}$ & $\num{0.21}$  & $\num{100}$  & $\num{3.30E-11}$ \\
    Medium      & $\num{8.59E+05}$ & $\num{18.42}$ & $\num{3000}$ & $\num{5.73E-05}$
                & $\num{8.54E+05}$ & $\num{0.17}$  & $\num{100}$  & $\num{1.77E-10}$ \\
    Fine        & $\num{1.68E+06}$ & $\num{23.40}$ & $\num{3000}$ & $\num{3.62E-05}$
                & $\num{1.39E+06}$ & $\num{0.34}$  & $\num{100}$  & $\num{1.39E-09}$ \\
    Very Fine   & $\num{4.99E+06}$ & $\num{37.80}$ & $\num{3000}$ & $\num{3.98E-05}$
                & $\num{3.55E+06}$ & $\num{0.71}$  & $\num{100}$  & $\num{1.02E-07}$ \\
    Body-Fitted & $\num{9.18E+05}$ & $\num{0.88}$  & $\num{4000}$ & $\num{9.36E-07}$
                & $\num{9.18E+05}$ & $\num{0.88}$  & $\num{4000}$ & $\num{9.36E-07}$ \\
    \arrayrulecolor{black}\hline
  \end{tabularx}
\end{table}
\subsection{Transonic flow around the RAE2822 airfoil}
\begin{figure}[h]
  \centering
  \includegraphics[width=0.49\linewidth]{\figurespath/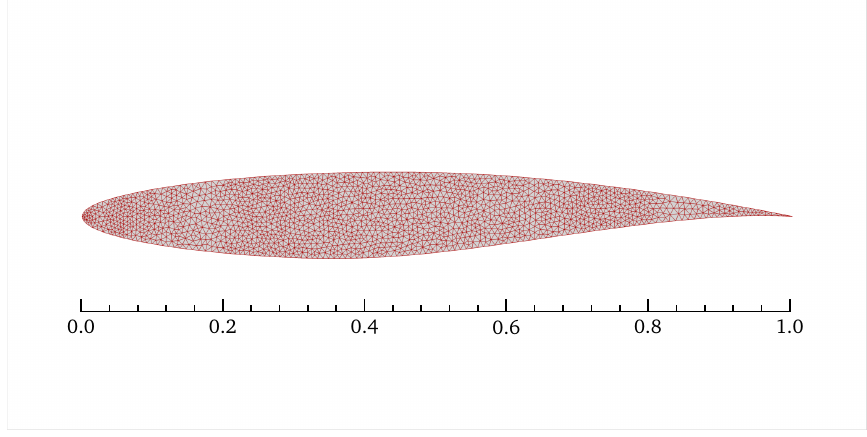}%
  \hfill%
  \includegraphics[width=0.49\linewidth]{\figurespath/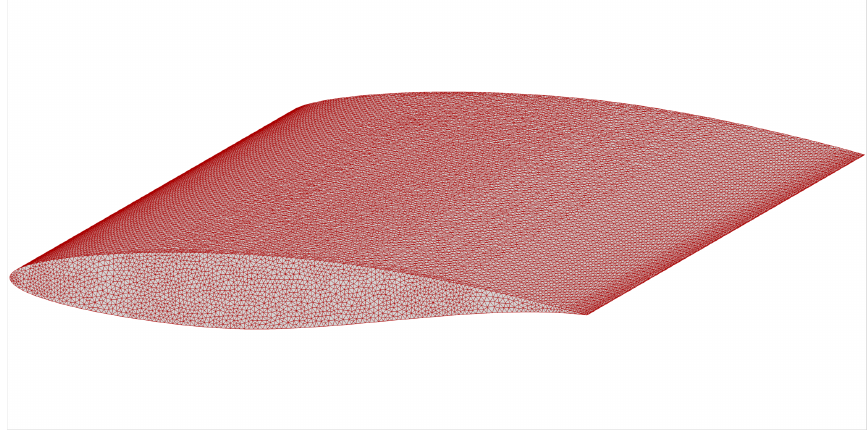}%
  \\[\medskipamount]
  \caption{Geometry used for the numerical simulations of the RAE2822 airfoil.}%
  \label{fig:numerical-computations:rae2822:geometry}%
\end{figure}

\begin{figure}[h]
  \centering
  \includegraphics[width=0.49\linewidth]{\figurespath/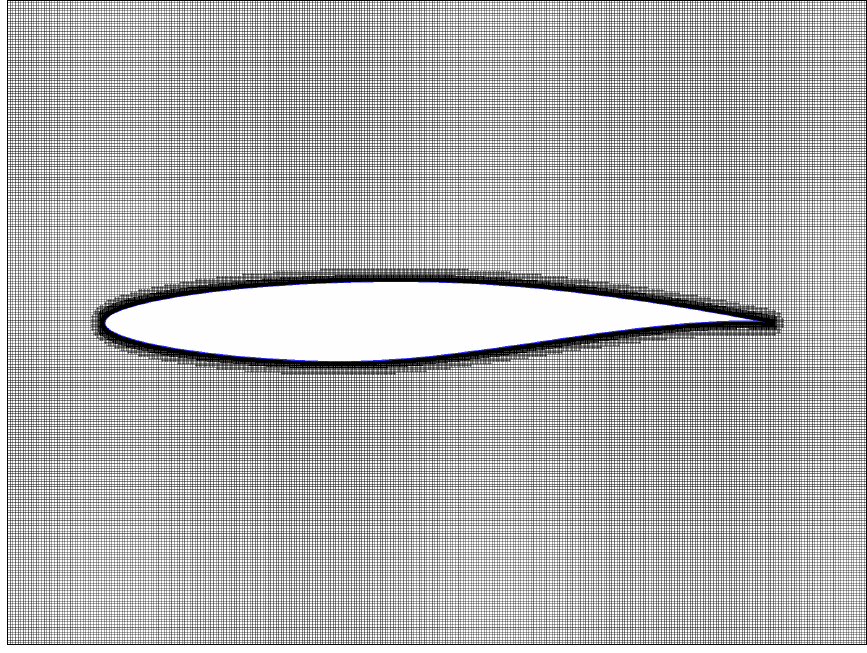}%
  \hfil%
  \includegraphics[width=0.49\linewidth]{\figurespath/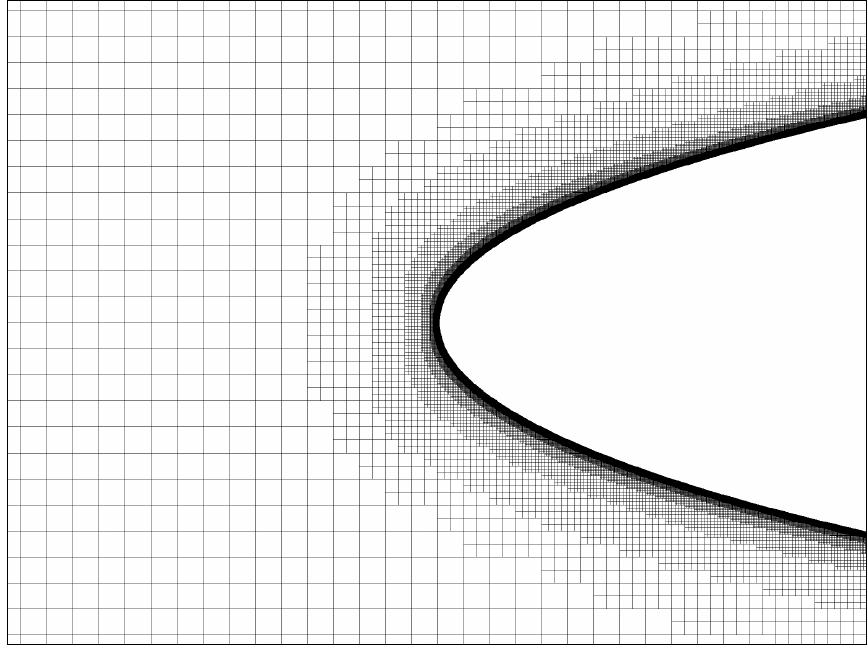}%
  \\[\medskipamount]
  \caption{Mesh refinement around the RAE2822 airfoil. The mesh has been generated with maximum level of refinement $l_{\mathrm{max}}=16$ and with the blanking activated. Close-up view of the mesh at $50x$ zoom, depicting the full airfoil (left) and close-up view of the mesh at $500x$ zoom, depicting the leading edge (right).}%
  \label{fig:numerical-computations:rae2822:mesh}%
\end{figure}

Now we will consider the transonic flow around the RAE2822 airfoil. This airfoil has been tested in the RAE wind tunnel at different flow conditions in the range of Mach numbers $\numrange[range-phrase=-]{0.676}{0.750}$ and at several Reynolds numbers. The test case considered in our study is the case $10$ from \cite{cook1979a}, which is representative of a transonic flow with a strong shock-boundary layer interaction with a re-attachment upstream of the trailing edge. The occurrence of a strong shock on the upper surface of the airfoil develops an important thickening of the boundary layer. Numerical and experimental data are available due to detailed studies carried out in the framework of the European initiative on validation of CFD codes (EUROVAL) project \cite{haase1993a}. Further numerical simulations have been performed in the context of RANS equations with finite volume methods in body-fitted meshes and also immersed boundary methods \cite{delanaye1999a,catalano2003a,capizzano2011a,capizzano2016a,constant2021a}. The geometry of the configuration is depicted in \cref{fig:numerical-computations:rae2822:geometry}.
The computational domain is similar to that used in the NACA0012 case, that is, a box with dimensions $[-20,+20]\times[-20,+20]\times[0,+1]$. The domain is initially discretized with $n_{x}n_{y}n_{z}=1600$ hexahedral elements, where $n_{x}=40$, $n_{y}=40$, and $n_{z}=1$. The mesh is then refined around the immersed geometry and in the wake region. The preprocessing tool refines the mesh around the immersed geometry up to the refinement level $l=16$, and the wake region is resolved with a refined subregion with refinement level $l=8$. This region is approximately $20$ airfoil chords downstream of the lead edge of the airfoil. In \cref{tab:numerical-computations:rae2822:mesh} are shown the characteristics of the meshes used for the mesh convergence study, ranging from a coarse mesh to a very fine mesh. The refinement level $l_{\mathrm{max}}$ listed in the table is the highest level of refinement around the immersed body.
\begin{table}[h]
  \setlength{\fboxsep}{0.00pt}%
  \renewcommand{\arraystretch}{1.2}%
  \newcolumntype{A}{X}%
  \newcolumntype{B}{p{0.2\textwidth}}%
  \caption{Characteristics of the meshes for the flow around the RAE2822 airfoil.}%
  \label{tab:numerical-computations:rae2822:mesh}%
  \begin{tabularx}{\textwidth}{AAAAX}
    \arrayrulecolor{black}\hline
    {Name} &
    {$l_{\mathrm{max}}$} &
    {$h_{\mathrm{wake}}$} &
    {$h_{\mathrm{body}}$} &
    {$\mathit{nElems}$} \\
    \arrayrulecolor{black}\hline
    Coarse      & $10$ & $\num{3.91E-03}$ & $\num{9.77E-04}$ & $\num{7.19E+05}$ \\
    Medium      & $12$ & $\num{3.91E-03}$ & $\num{2.44E-04}$ & $\num{8.54E+05}$ \\
    Fine        & $14$ & $\num{3.91E-03}$ & $\num{6.10E-05}$ & $\num{1.39E+06}$ \\
    Very Fine   & $16$ & $\num{3.91E-03}$ & $\num{1.53E-05}$ & $\num{3.55E+06}$ \\
    \arrayrulecolor{black}\hline
  \end{tabularx}
\end{table}
The immersed geometry is located in the center of the computational domain, with its cross-section in the plane $z$, and this corresponds to the RAE2822 profile with chord length $c=1.0$. The triangulation of the high-resolution surface geometry is made up of $\num{4744764}$ triangles. In \cref{fig:numerical-computations:rae2822:mesh} the mesh around the immersed geometry is depicted.\par
Regarding the flow conditions, the gas has adiabatic index $\gamma=1.4$ and it flows with Reynolds number $\mathrm{Re}=\num{6.5E6}$ and freestream Mach number $M=\num{0.729}$. The simulations were performed with the airfoil at an angle of attack $\alpha=\SI{2.31}{\degree}$. The non-dimensional density is set to $\rho=\num{1}$, and the non-dimensional pressure $p=\num{1}$. On the boundaries of the computational domain the following boundary conditions were set: the left face is set to inflow, the right face to outflow, the front and back faces were set to periodic, and the top and bottom to far field. The inflow pressure assumes the value of the stagnation pressure and the outflow pressure to the static pressure.\par
\begin{figure}[h]
  \centering
  \includegraphics[width=0.49\linewidth]{\figurespath/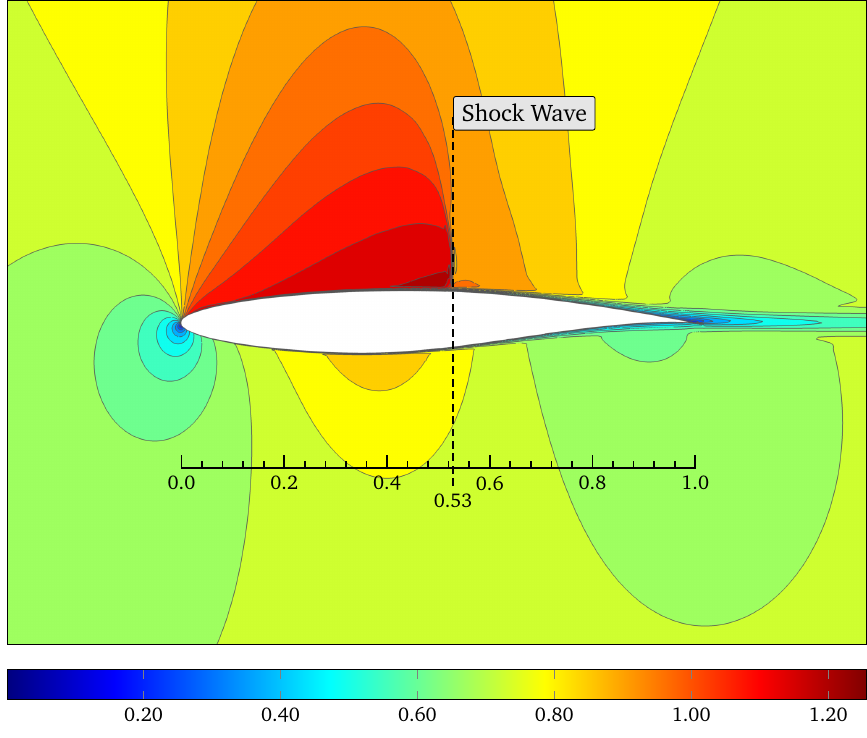}%
  \hfil%
  \includegraphics[width=0.49\linewidth]{\figurespath/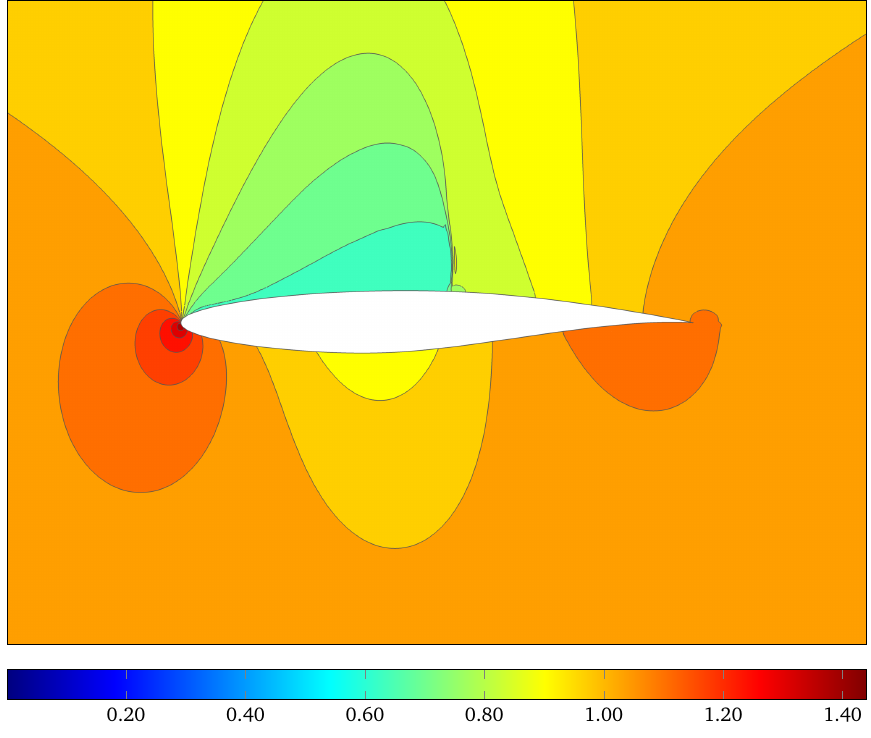}%
  \\[\medskipamount]
  \caption{Transonic flow around the RAE2822 airfoil. Contour plots of the Mach number (left) and pressure (right) for the simulations with angle of attack $\alpha=\SI{2.31}{\degree}$ and Mach number $M=\num{0.729}$. Computations were performed using the CODA solver based on an IBM Cartesian mesh with maximum level of refinement $l_{\mathrm{max}}=16$. The RANS equations were solved with the wall model deactivated.}%
  \label{fig:numerical-computations:rae2822:transonic:machnumber+pressure}%
\end{figure}
Contour plots of the Mach number and pressure from the numerical solution on the very fine mesh are shown in \cref{fig:numerical-computations:rae2822:transonic:machnumber+pressure} (left and right figures, respectively).
\begin{figure}[ht]
  \centering
  \includegraphics[width=0.49\linewidth]{\figurespath/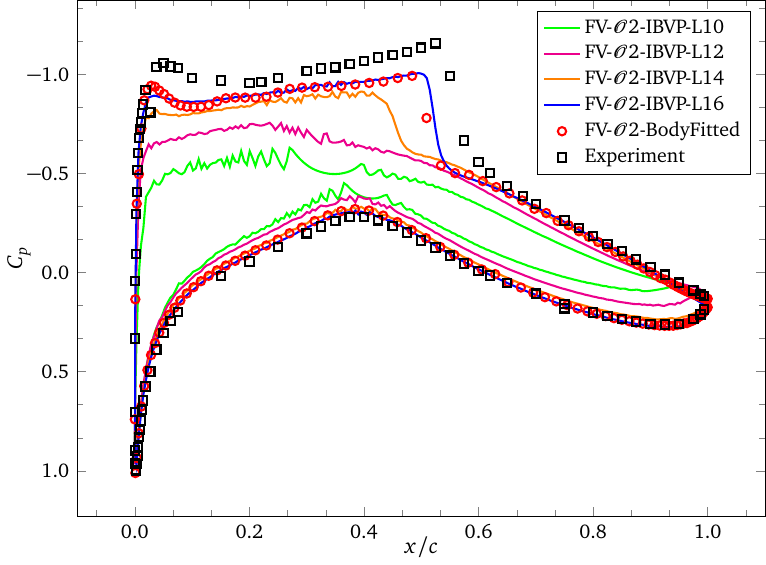}%
  \\[\medskipamount]
  \caption{Transonic flow around the RAE2822 airfoil. Grid convergence study in terms of the pressure coefficient $C_{p}$ for the transonic flow at angle of attack $\alpha=\SI{2.31}{\degree}$. Only results with wall model deactivated are shown.}%
  \label{fig:numerical-computations:rae2822:transonic:cp}%
\end{figure}
In the \cref{fig:numerical-computations:rae2822:transonic:cp} is presented the grid convergence in terms of the pressure coefficient $C_{p}$ on the airfoil surface, with wall model deactivated. The experimental data has been taken from \cite{cook1979a}. On the upper surface, the results show good agreement for the finer mesh with the body fitted simulation. Let us remind the reader that these results were obtained without including any penalization in the continuity equation, as proposed by other authors \cite{menez2023a}.
\subsection{Subsonic flow around the MDA30P30N multi-element airfoil}
\begin{figure}[h]
  \centering
  \includegraphics[width=0.49\linewidth]{\figurespath/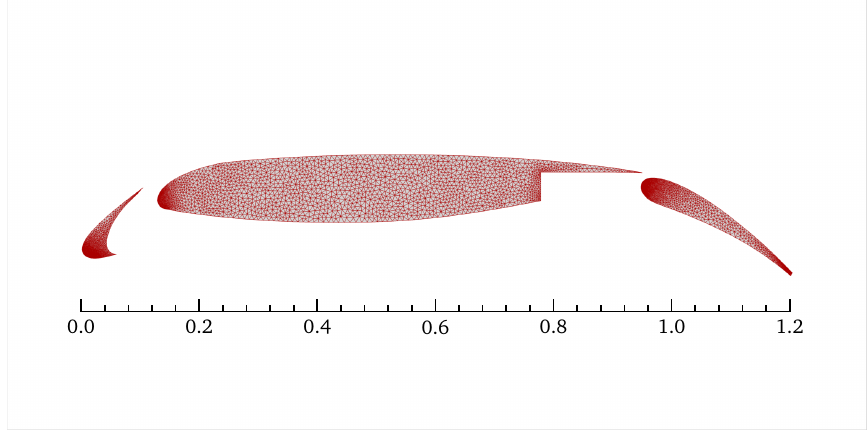}%
  \hfill%
  \includegraphics[width=0.49\linewidth]{\figurespath/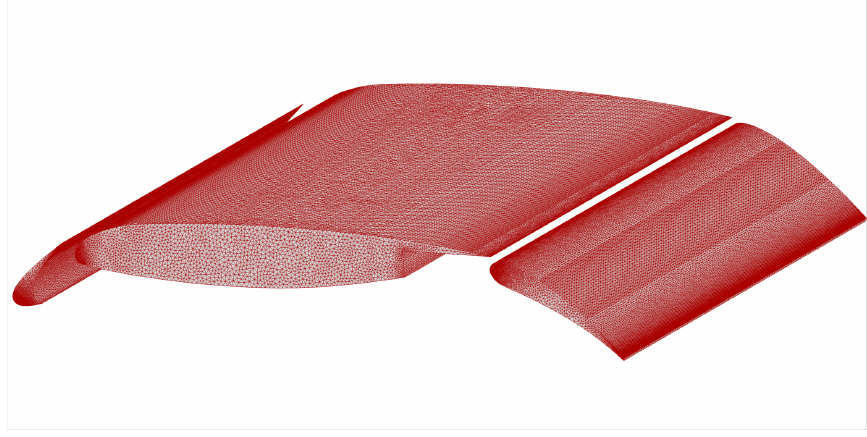}%
  \\[\medskipamount]
  \caption{Geometry used for the numerical simulations of the MDA30P30N multi-element airfoil.}%
  \label{fig:numerical-computations:mda30p30n:geometry}%
\end{figure}

The fourth test in the benchmarking is a three-element configuration, specifically the McDonnell Douglas 30P-30N landing configuration (MDA30P30N). Many experimental and computational studies have been performed for the flow past multi-element airfoils in the last decades \cite{valarezo1991a,valarezo1992a,chin1993a,rogers1994a,rogers1994b,anderson1995a,klausmeyer1997a,bertelrud1998a,rumsey1998a,liou1999a,spaid2000a}. Accurate prediction of the flow over multi-element airfoils during high-lift operations can improve the performance and the safety factor of aircrafts. The flow around multi-element airfoils is complex, and it is well known to be dominated by different flow mechanisms at different operating conditions, making rather difficult to accurately predict high-lift flow fields. The geometry of the configuration is depicted in \cref{fig:numerical-computations:mda30p30n:geometry}. The leading edge slat and the trailing edge flap have a deflection angle of $\SI{30}{\degree}$. This airfoil has been extensively tested in the NASA Langley Low Turbulence Pressure Tunnel at various Reynolds and Mach numbers and has also been numerically simulated using a wide range of numerical techniques to solve the Navier--Stokes equations along with different turbulence models.\par
\begin{figure}[h]
  \centering
  \includegraphics[width=0.49\linewidth]{\figurespath/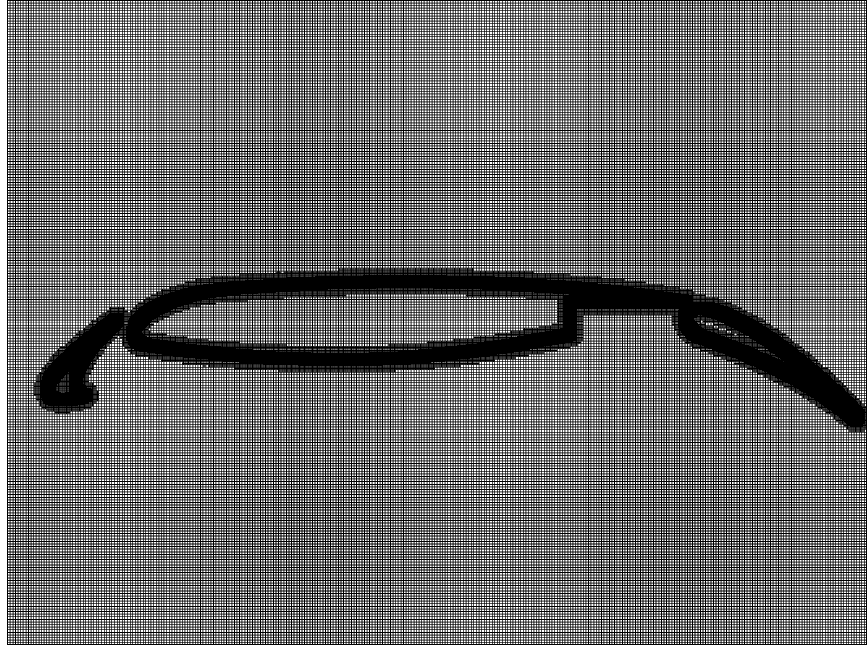}%
  \hfil%
  \includegraphics[width=0.49\linewidth]{\figurespath/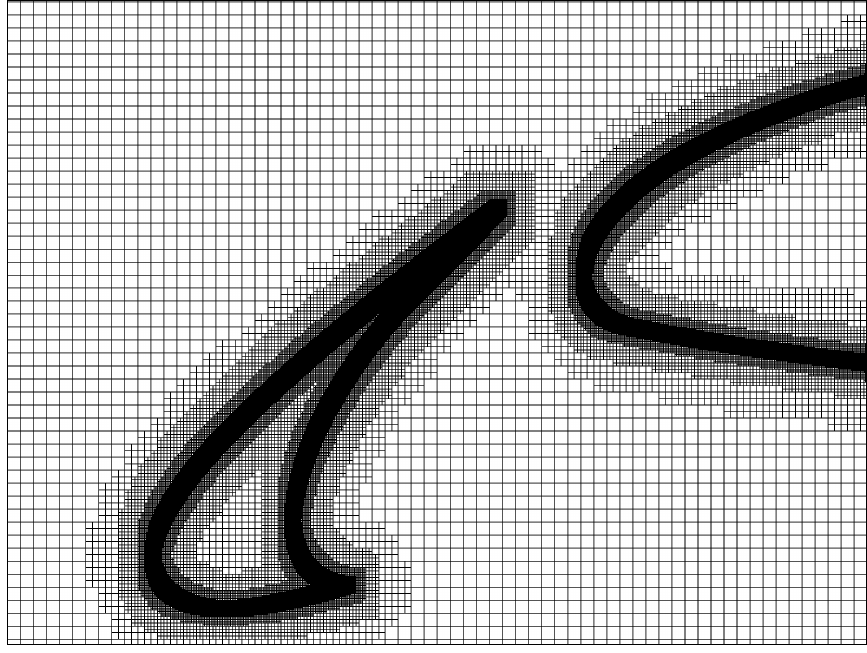}%
  \\[\medskipamount]
  \caption{Mesh refinement around the MDA30P30N airfoil. The mesh has been generated with maximum level of refinement $l_{\mathrm{max}}=16$ and with the blanking deactivated. Close-up view of the mesh at $50x$ zoom, depicting the full airfoil (left) and close-up view of the mesh at $500x$ zoom, depicting the slat and leading edge (right).}%
  \label{fig:numerical-computations:mda30p30n:mesh}%
\end{figure}

The computational domain is similar to that used in the NACA0012 case, that is, a box with dimensions $[-20,+20]\times[-20,+20]\times[0,+1]$. The domain is initially discretized with $n_{x}n_{y}n_{z}=1600$ hexahedral elements, where $n_{x}=40$, $n_{y}=40$, and $n_{z}=1$. The mesh is then refined around the immersed geometry and in the wake region. The mesh around the immersed geometry is refined to the refinement level $l=16$, and the wake region is resolved with a refined subregion with refinement level $l=8$. This region is approximately 20 airfoil chords downstream of the lead edge of the airfoil. In \cref{tab:numerical-computations:mda30p30n:mesh} are shown the characteristics of the meshes used for the mesh convergence study, ranging from a medium mesh to a very fine mesh. The refinement level $l_{\mathrm{max}}$ listed in the table is the highest level of refinement around the immersed body.
\begin{table}[h]
  \setlength{\fboxsep}{0.00pt}%
  \renewcommand{\arraystretch}{1.2}%
  \newcolumntype{A}{X}%
  \newcolumntype{B}{p{0.2\textwidth}}%
  \caption{Characteristics of the meshes for the flow around the MDA30P30N multi-element airfoil.}%
  \label{tab:numerical-computations:mda30p30n:mesh}%
  \begin{tabularx}{\textwidth}{AAAAX}
    \arrayrulecolor{black}\hline
    {Name} &
    {$l_{\mathrm{max}}$} &
    {$h_{\mathrm{wake}}$} &
    {$h_{\mathrm{body}}$} &
    {$\mathit{nElems}$} \\
    \arrayrulecolor{black}\hline
    Coarse      & $10$ & $\num{3.91E-03}$ & $\num{9.77E-04}$ & $\num{9.64E+05}$ \\
    Medium      & $12$ & $\num{3.91E-03}$ & $\num{2.44E-04}$ & $\num{1.25E+06}$ \\
    Fine        & $14$ & $\num{3.91E-03}$ & $\num{6.10E-05}$ & $\num{2.44E+06}$ \\
    Very Fine   & $16$ & $\num{3.91E-03}$ & $\num{1.53E-05}$ & $\num{7.19E+06}$ \\
    \arrayrulecolor{black}\hline
  \end{tabularx}
\end{table}
The immersed geometry is located in the center of the computational domain, with its cross-section in the plane $z$, and this corresponds to the MDA30P30N profile with chord length $c=1.2$. The triangulation of the surface geometry is made up of $\num{6381338}$ triangles. In \cref{fig:numerical-computations:mda30p30n:mesh} the mesh around the immersed geometry is depicted. The inner elements are not shown.\par
Regarding flow conditions, the gas has an adiabatic index $\gamma=1.4$ and flows with Reynolds number $\mathrm{Re}=\num{9E6}$ and freestream Mach number $M=\num{0.2}$. The simulations were performed with the airfoil at an angle of attack $\alpha=\SI{8}{\degree}$. The non-dimensional density is set to $\rho=\num{1}$, and the non-dimensional pressure $p=\num{1}$. On the boundaries of the computational domain the following boundary conditions were set: the left face is set to inflow, the right face to outflow, the front and back faces were set to periodic, and the top and bottom to far field. The inflow pressure assumes the value of the stagnation pressure and the outflow pressure to the static pressure.\par
\begin{figure}[h]
  \centering
  \includegraphics[width=0.49\linewidth]{\figurespath/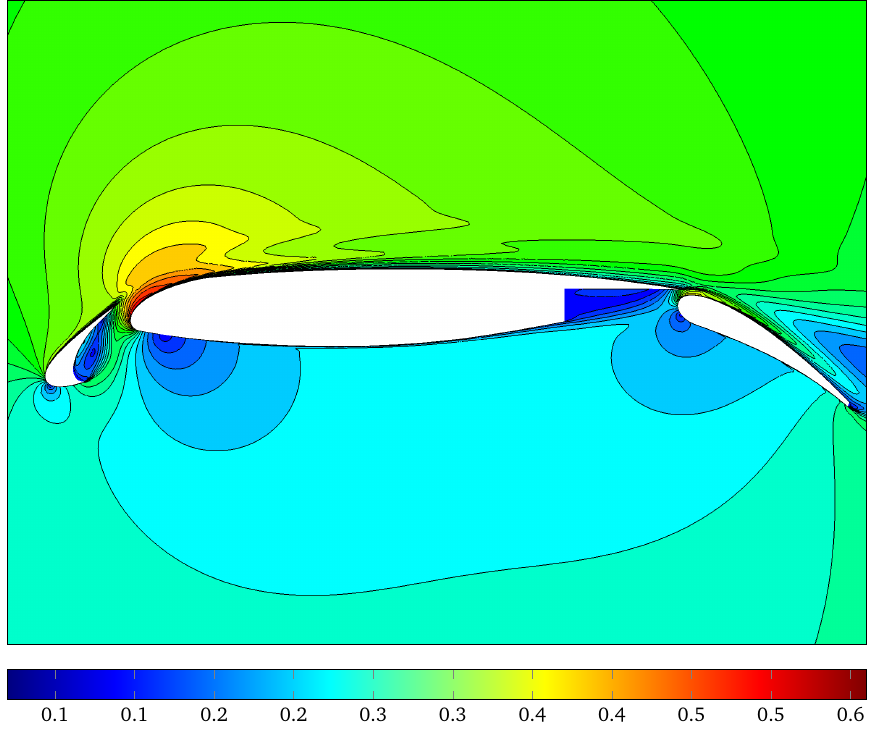}%
  \hfil%
  \includegraphics[width=0.49\linewidth]{\figurespath/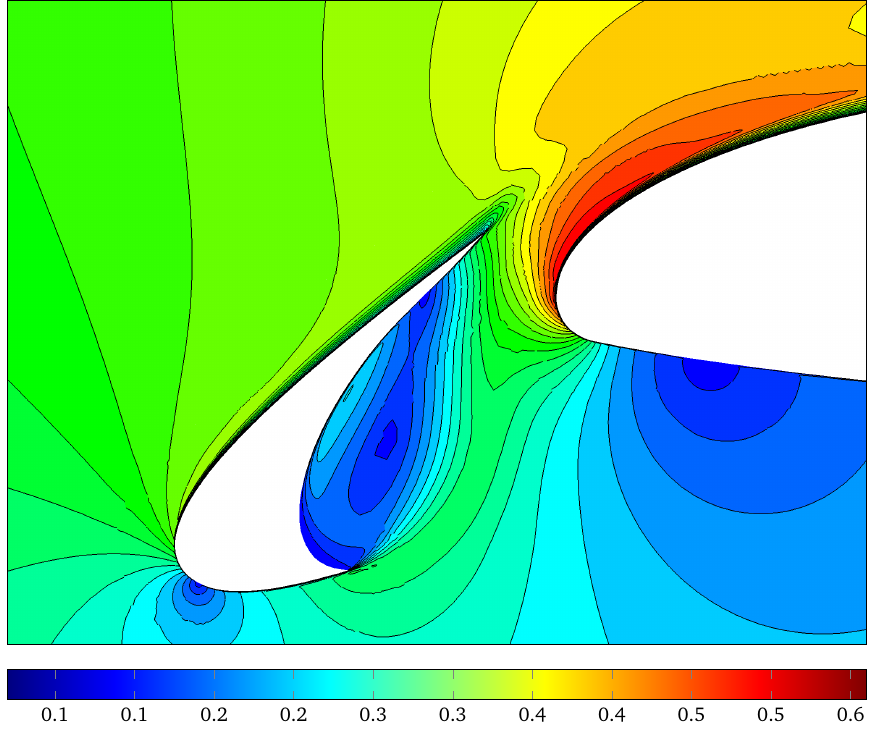}%
  \\[\medskipamount]
  \caption{Subsonic flow around the MDA30P30N multi-element airfoil. Contour plots of the velocity magnitude for the simulations with angle of attack $\alpha=\SI{8}{\degree}$. Computations were performed using the CODA solver based on an IBM Cartesian mesh with a level of refinement around the geometry $l_{\mathrm{max}}=16$. The RANS equations were solved with the wall model deactivated.}%
  \label{fig:numerical-computations:mda30p30n:subsonic:velocity}%
\end{figure}
Contour plots of velocity magnitude from the numerical solution on the very fine mesh are shown in \cref{fig:numerical-computations:mda30p30n:subsonic:velocity}. A small separation region is present on the suction side of the upstream side of the trailing edge of the flap.
\begin{figure}[ht]
  \centering
  \includegraphics[width=0.49\linewidth]{\figurespath/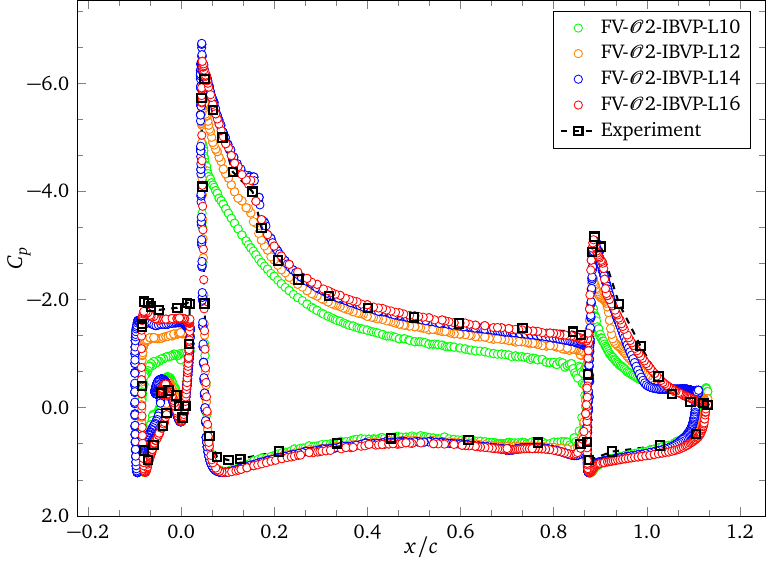}%
  \\[\medskipamount]
  \caption{Subsonic flow around the MDA30P30N multi-element airfoil. Grid convergence study in terms of the pressure coefficient $C_{p}$ for the subsonic flow at angle of attack $\alpha=\SI{8}{\degree}$. Only results with wall model deactivated are shown.}%
  \label{fig:numerical-computations:mda30p30n:subsonic:cp-cf}%
\end{figure}
In \cref{fig:numerical-computations:mda30p30n:subsonic:cp-cf} is presented the grid convergence in terms of the pressure coefficient $C_{p}$ on the multi-element airfoil surface, with wall model deactivated. The experimental data were taken from \cite{murayama2014b}. In the main part of the airfoil, the pressure coefficient curve agrees well with the experimental data, except at its trailing edge. We can observe a remarkable discrepancy with respect to the experimental data on the upper surface of the slat and the flap for the coarse and medium meshes, but a good agreement is obtained in the simulations with the fine and very fine meshes. We observe that the surface skin friction coefficient is not very well predicted, and presents an oscillatory behavior.
The lift and drag values are shown in \cref{tab:numerical-computations:mda30p30n:subsonic:aoa-08:lift-drag} for the angle of attack $\alpha=\SI{8}{\degree}$. The lift coefficient increases monotonically as the grid is refined, getting closer to the experimental value $C_{L}=\num{3.243}$ \cite{murayama2014b}. No experimental data is available for the drag coefficient.
\begin{table}[h]
  \setlength{\fboxsep}{0.00pt}%
  \renewcommand{\arraystretch}{1.2}%
  \newcolumntype{A}{C}%
  \newcolumntype{B}{p{0.2\textwidth}}%
  \caption{Subsonic flow around the MDA30P30N multi-element airfoil. Lift and drag coefficients for angle of attack $\alpha=\SI{8}{\degree}$.}%
  \label{tab:numerical-computations:mda30p30n:subsonic:aoa-08:lift-drag}%
  \begin{tabularx}{\textwidth}{XAA}
    \arrayrulecolor{black}\hline
    Mesh &
    $C_{L}$ &
    $C_{D}$ \\
    \arrayrulecolor{black}\hline
    Coarse       & $\num{2.473}$ & $\num{4.490E-02}$ \\
    Medium       & $\num{2.743}$ & $\num{4.579E-02}$ \\
    Fine         & $\num{2.837}$ & $\num{4.673E-02}$ \\
    Very Fine    & $\num{3.017}$ & $\num{5.851E-02}$ \\
    Experimental & $\num{3.243}$ & Not Available \\
    \arrayrulecolor{black}\hline
  \end{tabularx}
\end{table}
\subsection{Subsonic flow around the NASA high-lift CRM}
The last test we consider in this work is the NASA high-lift CRM configuration, selected from the Fourth AIAA CFD High Lift Prediction Workshop \cite{ashton2024a}. The NASA Common Research Model has been used mainly in the Drag Prediction Workshop (DPW) and the High Lift Prediction Workshop (HLPW) to assess the accuracy of numerical methods in the prediction of aircraft forces and moments. The database of these workshops allows us to assess the robustness and effectiveness of state-of-the-art numerical programs and turbulence modeling techniques using Navier--Stokes solvers.\par
The CRM configuration was originally designed by Boeing and was further manufactured and tested by NASA \cite{vassberg2008a}.
The NASA CRM consists of a contemporary supercritical transonic wing and a fuselage that is representative of a wide-body commercial transport aircraft. Several CRM experiments have been carried out in several research facilities \cite{rivers2014a,ueno2015a,cartieri2018a}.\par
\begin{figure}[h]
  \centering
  \includegraphics[width=0.75\linewidth]{\figurespath/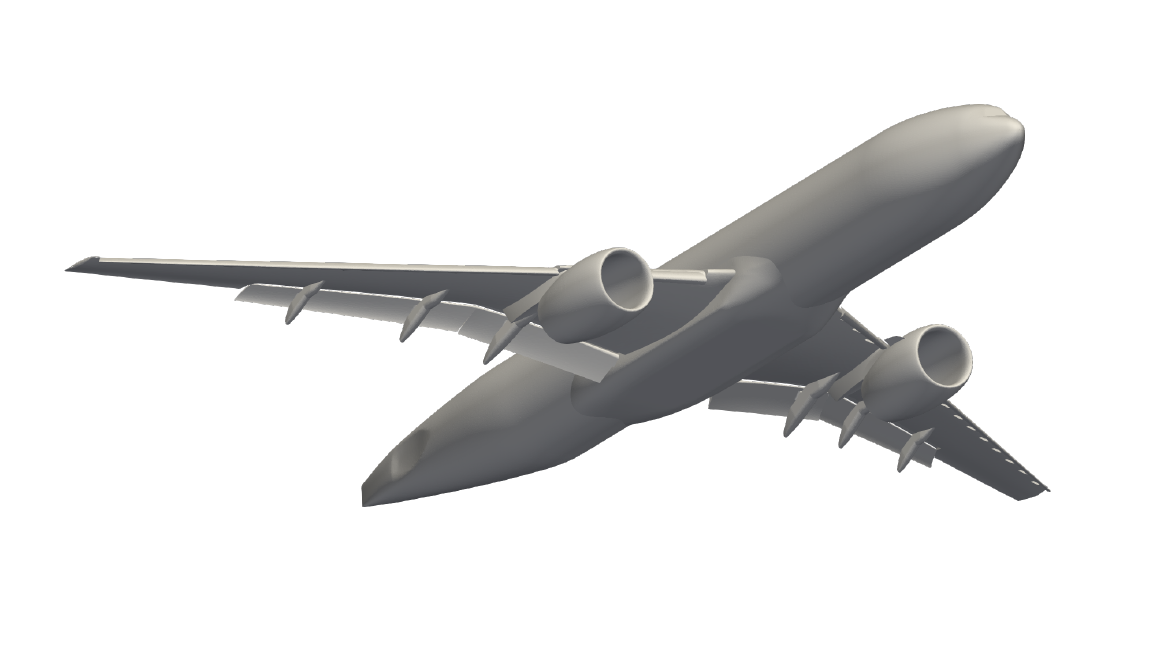}%
  \\[\medskipamount]
  \caption{Geometry used for the numerical simulations of the NASA High-Lift CRM Wing-Body-Flap-Slat-Nacelle-Pylon.}%
  \label{fig:numerical-computations:crm-wbfsnp:geometry}%
\end{figure}
\begin{figure}[h]
  \centering
  \includegraphics[width=0.49\linewidth]{\figurespath/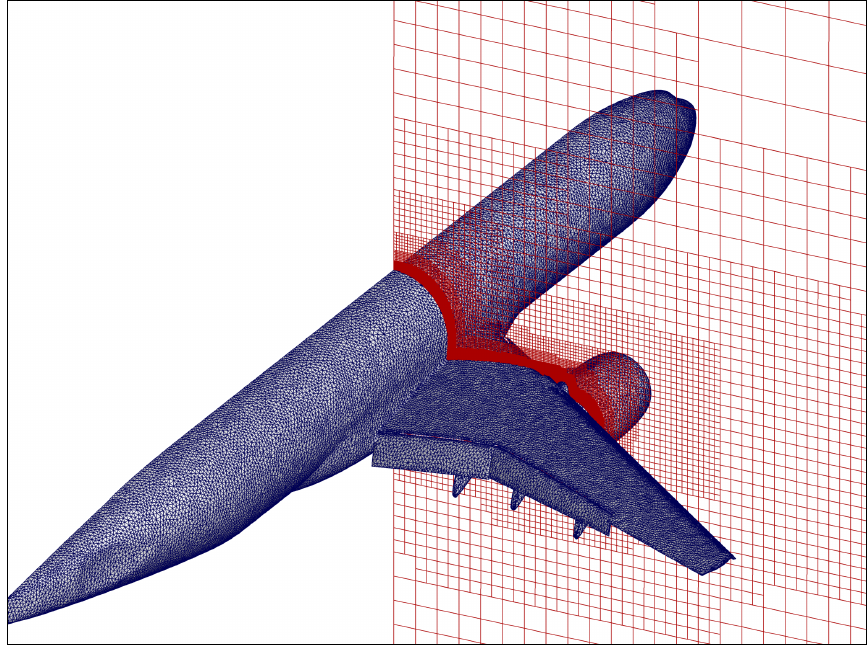}%
  \hfil%
  \includegraphics[width=0.49\linewidth]{\figurespath/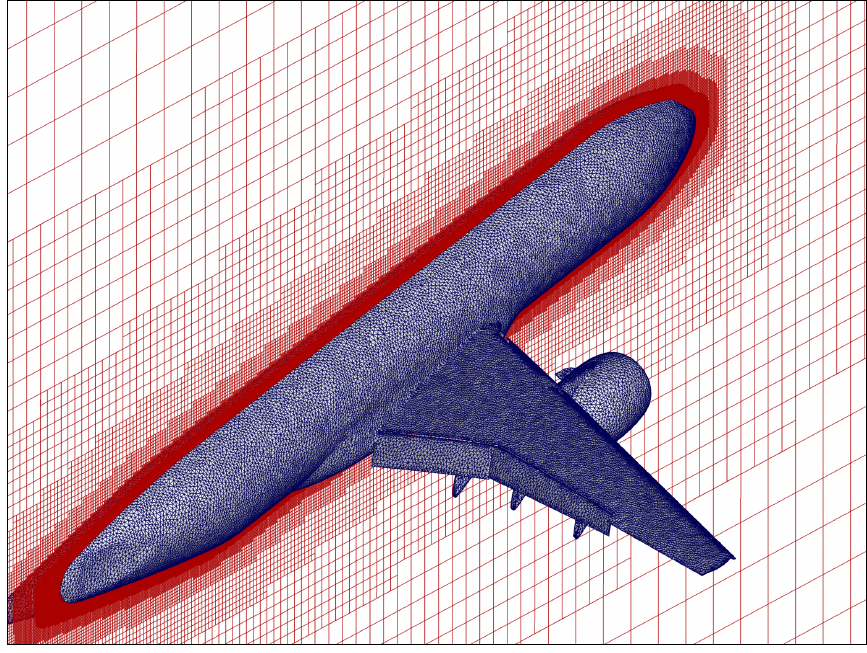}%
  \\[\medskipamount]
  \caption{Mesh refinement around the NASA high-lift CRM Wing-Body-Flap-Slat-Nacelle-Pylon geometry.}%
  \label{fig:numerical-computations:crm-wbfsnp:mesh}%
\end{figure}

The computational domain used for this simulation is a box with dimensions $[-20,+20]\times[-20,+20]\times[0,+20]$. The domain is initially discretized with $n_{x}n_{y}n_{z}=\num{32000}$ hexahedral elements, where $n_{x}=40$, $n_{y}=40$, and $n_{z}=20$. The mesh is then refined around the immersed geometry and in the wake region. The mesh around the immersed geometry is refined to the refinement level $l=10$, and the wake region is resolved with a refined subregion with refinement level $l=8$. In \cref{tab:numerical-computations:crm-wbfsnp:mesh} are shown the characteristics of the meshes used for the mesh convergence study, ranging from a very coarse mesh to a medium mesh. The refinement level $l_{\mathrm{max}}$ listed in the table is the highest level of refinement around the immersed geometry. The geometry of the configuration is depicted in \cref{fig:numerical-computations:crm-wbfsnp:geometry}.
\begin{table}[h]
  \setlength{\fboxsep}{0.00pt}%
  \renewcommand{\arraystretch}{1.2}%
  \newcolumntype{A}{X}%
  \newcolumntype{B}{p{0.2\textwidth}}%
  \caption{Characteristics of the meshes for the flow around the NASA high-lift CRM Wing-Body-Flap-Slat-Nacelle-Pylon geometry.}%
  \label{tab:numerical-computations:crm-wbfsnp:mesh}%
  \begin{tabularx}{\textwidth}{AAAAX}
    \arrayrulecolor{black}\hline
    {Name} &
    {$l_{\mathrm{max}}$} &
    {$h_{\mathrm{wake}}$} &
    {$h_{\mathrm{body}}$} &
    {$\mathit{nElems}$} \\
    \arrayrulecolor{black}\hline
    Very Coarse & $8$  & $\num{3.91E-03}$ & $\num{3.91E-03}$ & $\num{5.827E+06}$ \\
    Coarse      & $9$  & $\num{3.91E-03}$ & $\num{1.95E-03}$ & $\num{23.207E+06}$ \\
    Medium      & $10$ & $\num{3.91E-03}$ & $\num{9.77E-04}$ & $\num{96.520E+06}$ \\
    \arrayrulecolor{black}\hline
  \end{tabularx}
\end{table}
The immersed geometry is located in the center of the computational domain. The triangulation of the surface geometry is made up of $\num{74331}$ triangles. This low-resolution tessellation will have a negative impact on the final numerical solution, as it could be observed in the computations of NACA0012 airfoil. In \cref{fig:numerical-computations:crm-wbfsnp:mesh} is depicted the mesh around the immersed geometry.\par
The Reynolds number based on the mean aerodynamic chord ($\mathrm{MAC}=\SI{7.005}{\metre}$) is $\mathrm{Re}=\num{5.6}$ million, and the angle of attack is $\alpha=\SI{8}{\degree}$ with an incoming Mach number of $M_{\infty}=\num{0.2}$. On the boundaries of the computational domain the following boundary conditions were set: the left face is set to inflow, the right face to outflow, the front and back faces were set to periodic, and the top and bottom to far field. The inflow pressure assumes the value of the stagnation pressure and the outflow pressure to the static pressure.\par
\begin{figure}[ht]
  \centering
  \includegraphics[width=0.49\linewidth]{\figurespath/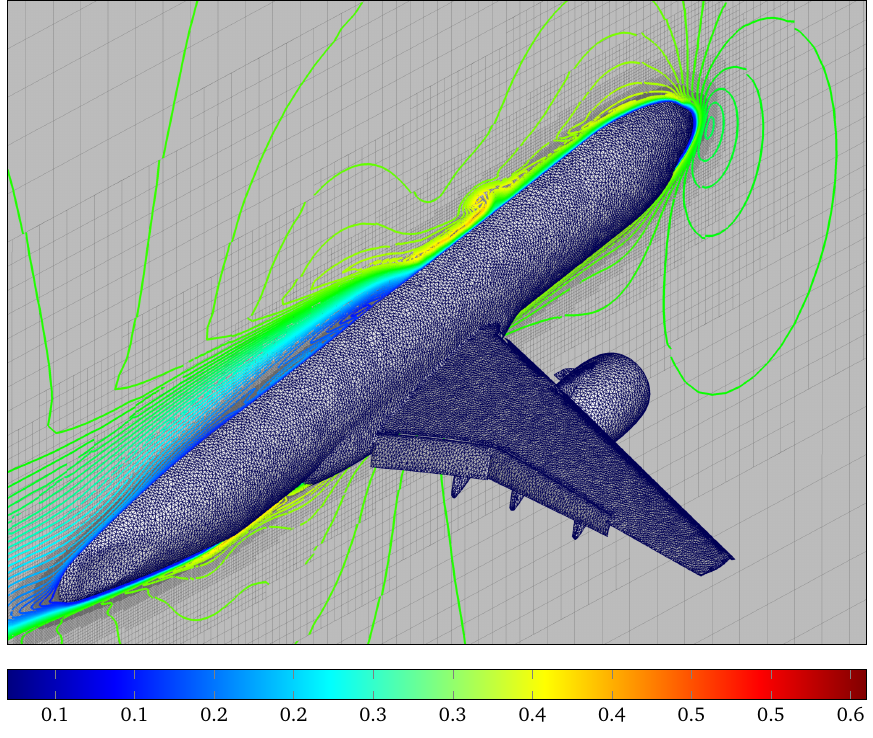}%
  \hfil%
  \includegraphics[width=0.49\linewidth]{\figurespath/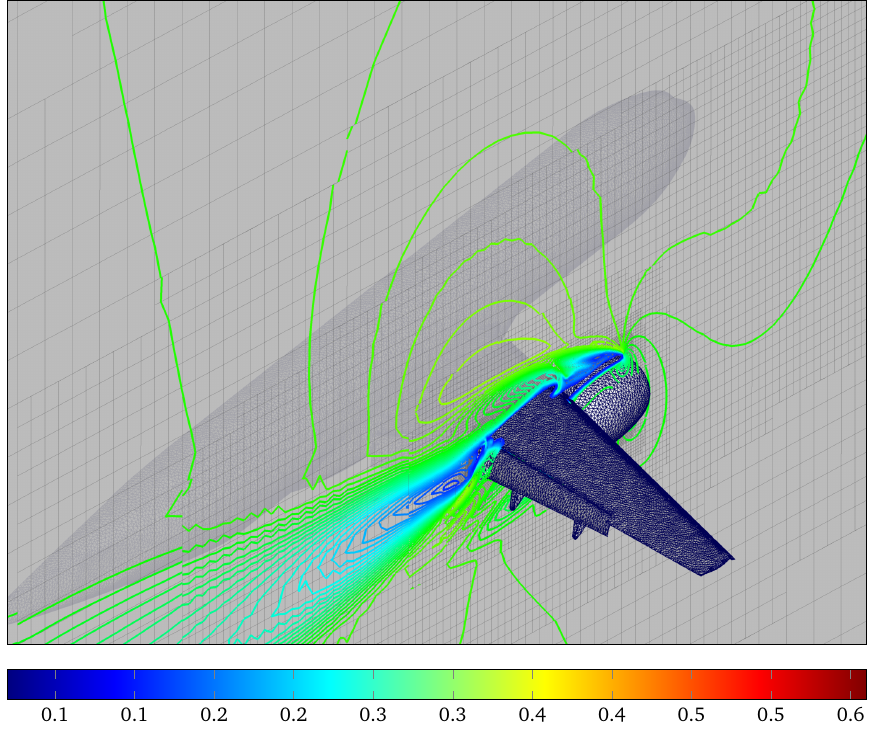}%
  \\[\medskipamount]
  \caption{Subsonic flow around the NASA high-lift CRM. Contour plots of the velocity magnitude at different positions for the simulations with angle of attack $\alpha=\SI{8}{\degree}$. Computations were performed using the CODA solver based on an IBM Cartesian mesh with a level of refinement around the geometry $l_{\mathrm{max}}=\num{10}$. The RANS equations were solved with the wall model deactivated.}%
  \label{fig:numerical-computations:crm-hl-wbfsnp:subsonic:velocity}%
\end{figure}
Contour plots of the velocity magnitude at different positions are shown in \cref{fig:numerical-computations:crm-hl-wbfsnp:subsonic:velocity}. The curves are qualitatively similar to those obtained in body-fitted simulations we have previously done with the solver CODA (not shown in this work). We can clearly observe flow separation on the surface of the fuselage (left) and on the suction side of the upstream side of the trailing edge of the flap.\par
Some final comments are necessary to mention at this point. In the computations of NACA0012, RAE2822 and MDA30P30N multi-element airfoils, we found that, in order to accurately capture the aerodynamics and obtain very good agreement with reference body-fitted simulations and experimental data, high-resolution surface representation of immersed geometries and very fine meshes around these are required. In a good IBM-Cartesian mesh, the element size close to the geometry is around $h_{\mathrm{body}}=\num{1.53E-05}$ ($y^{+}\approx{100}$), or even smaller. For the case of the NASA high-lift CRM, a mesh with those characteristics would have a very high cell number, typically over several billions. Cartesian meshes with up to $\num{300}$ million elements have been generated in a fast manner with our preprocessing tool in $\num{4}$ hours on a single core machine, but its generation requires around $\num{2}$ terabytes of RAM. Besides, the simulations in CODA will need thousands of cores and several terabytes of RAM if such meshes are employed. With our limited computational resources, we were able only to simulate the NASA high-lift CRM on a medium-size mesh with $\num{97}$ million cells and with element size close to the geometry $h_{\mathrm{body}}=\num{9.77E-04}$. Taking into account all these previous considerations, in our computations of the NASA high-lift CRM on the medium-size mesh, this lack of resolution around the immersed body had as a consequence poor quality in the integral quantities like lift and drag coefficients, and the pressure and skin friction curves, this means that they are not in good agreement with experimental data, but this issue can be solved by using finer meshes. This is consistent with the results shown before for the MDA30P30N airfoil. We hope that all previous sections have already proved the validity of the implementation. This last section aims only to show qualitative results (not quantitative) to highlight the potential of the methodology for more general complex cases.
%
%
\section{Conclusions}\label{sec:conclusions}
The immersed boundary volume penalization method for the Navier--Stokes and the RANS equations has been implemented in the CFD solver CODA. The immersed boundary method is compatible with the finite volume, modal discontinuous Galerkin and discontinuous Galerkin spectral element methods and with the explicit and implicit temporal schemes available in CODA. An efficient preprocessing tool for the construction of unstructured hexahedral meshes with adaptive mesh refinement around immersed geometries has been developed. The octree meshes generated with the tool are imported in CODA and used for simulating complex aerodynamics flow problems. Immersed boundary methods require a very simple meshing process: only the box dimension, initial number of elements of the Cartesian mesh, maximum level of refinement, and the geometry of the body are required as input. The automatic mesh generation process is able to automatically refine the mesh near the body. Besides, the immersed boundary technology can be used straightforward with an external predefined mesh, without any additional modification.\par
An important aspect regarding the volume penalization methodology is that the use of source terms increases the stiffness of the partial differential equations, and therefore, more iterations are required until convergence. This represents a fundamental disadvantage with respect to CFD solvers based on body-fitted meshes and also other flavors of immersed boundary methods. Mesh blanking in IBM Cartesian meshes aims to reduce the computation time when similar results are sought among CFD solvers based on IBM techniques and body-fitted meshes. We have explored the mesh blanking to assess the performance of the CODA solver with IBM. Computations done using blanked meshes have a superior efficiency compared to simulations using meshes without blanking. We have found that simulations using blanked meshes need up to two orders of magnitude fewer linear iterations than simulations with no blanked meshes in order to converge to a solution with similar residual in the density in computations performed with body-fitted meshes.\par
Several numerical computations were performed and discussed: subsonic flow past the NACA0012 airfoil, transonic flow past the RAE2822 airfoil, subsonic flow past the MDA30P30N multi-element airfoil and subsonic flow around the NASA high-lift CRM aircraft. The numerical computations for the airfoils are in good agreement with their corresponding experimental data for the pressure coefficient curve. The skin friction curves have an oscillatory behavior for low resolution meshes, but these oscillations tend to decrease once the geometry and mesh are refined. Drag and lift coefficients have values very close to the reference data, but our findings allow us to conclude that even finer meshes are required if an excellent agreement is sought. Computations with deactivation and activation of the wall model are quite similar, but computations are more robust with deactivated wall model. More research is necessary to improve the robustness of the method.\par
Even though the immersed boundary methodology works within the CFD solver CODA with very good results, a critical disadvantage of these techniques with respect to a body-fitted approach is that the mesh refinement around the immersed geometry leads to a very high number of elements when a high resolution of the boundary layer is desired. Anisotropic mesh refinement reduce the number of elements in comparison with isotropic mesh refinement, but the mesh size is still prohibitive.\par
As a consequence of these considerations, we will explore also alternative approaches to automatically generate meshes for IBM simulations. In particular, future work will focus on hybrid IBM approaches, in which only geometrical details (control surfaces, ice shapes, etc) will be immersed in a body-fitted, wall-resolved mesh of a clean aerodynamic surface. This approach would benefit from IBM advantages (several configurations of the geometrical details can be tested on the same body-fitted background mesh) without handling meshes of very big sizes.
%
%
\printcredits
%
%
\section{Acknowledgments}
Jonatan Núñez, David Huergo, Esteban Ferrer and Eusebio Valero acknowledge the funding received by the Grant NextSim\slash AEI\slash 10.13039\slash 501100011033 and H2020, GA-956104. This project has received funding from the Clean Aviation Joint Undertaking under the European Union’s Horizon Europe research and innovation programme under Grant Agreement HERA (Hybrid-Electric Regional Architecture) no. 101102007. Views and opinions expressed are, however, those of the author(s) only and do not necessarily reflect those of the European Union or CAJU. Neither the European Union nor the granting authority can be held responsible for them. Esteban Ferrer and Eusebio Valero acknowledge the funding received by the Grant DeepCFD (Project No. PID2022-137899OB-I00) funded by MCIN\slash AEI\slash 10.13039\slash 501100011033 and by ERDF A way of making Europe. Esteban Ferrer would like to thank the support of the Comunidad de Madrid and Universidad Politécnica de Madrid for the Young Investigators award: APOYO-JOVENES-21-53NYUB-19-RRX1A0. Finally, all authors gratefully acknowledge Universidad Politécnica de Madrid (\url{www.upm.es}) for providing computing resources on Magerit Supercomputer. The authors also thankfully acknowledge the computer resources at MareNostrum and the technical support provided by Barcelona Supercomputing Center (RES-IM-2022-3-0023).
%
%

%
%
%
\end{document}